\documentclass[10pt,journal,letterpaper,compsoc]{IEEEtran}

\IEEEoverridecommandlockouts

\usepackage{graphicx}
\usepackage{url}
\usepackage{stmaryrd}
\usepackage{amssymb,amsmath}
\usepackage{multirow,url}
\usepackage{color,cite}

\newtheorem{lemma}{Lemma}

\newtheorem{proposition}{Proposition}

\newtheorem{theorem}{Theorem}
\newtheorem{observation}{Observation}

\ifodd 1

\newcommand{\com}[1]{\textbf{\color{red}(COMMENT: #1)}} 
\newcommand{\comg}[1]{\textbf{\color{green} (COMMENT: #1)}}
\newcommand{\response}[1]{\textbf{\color{magenta} (RESPONSE: #1)}} 
\newcommand{\del}[1]{}
\else

\newcommand{\com}[1]{}
\newcommand{\comg}[1]{}
\newcommand{\response}[1]{}
\newcommand{\del}[1]{}
\fi

\begin{document}

\title{Pricing for Local and Global WiFi Markets}

\IEEEcompsoctitleabstractindextext{%
\begin{abstract}

This paper analyzes two pricing schemes commonly used in WiFi markets: the flat-rate and the usage-based pricing. The flat-rate pricing encourages {{the} maximum usage}, while {the usage-based pricing can flexibly attract more users especially those with low valuations in mobile Internet access.} First, we use theoretical analysis to compare the two schemes and show that for a single provider in a market, as long as the WiFi capacity is abundant, the flat-rate pricing leads to more revenue. Second, we study how a global provider (e.g., Skype) {collaborates with this monopolist in each local market} to provide a global WiFi service. {We} formulate {the interactions between the global and local providers as} a dynamic game. In Stage I, {the global provider bargains with the local provider in each market {to determine} the global WiFi service price and revenue sharing agreement}. In Stage II, local users and travelers choose local or global WiFi services. {We analytically show that the global provider prefers to use the usage-based pricing to avoid a severe competition with the local provider.} At the equilibrium, the global provider always shares the majority of his revenue {with the local provider} to {incentivize the cooperation}. Finally, {we analytically study how the interaction changes if the local market has more than one local provider. In this case, the global provider can integrate the coverages of multiple local providers and provide a better service. Compared to the local monopoly case, local market competition enables the global provider to share less revenue with each of the local providers.} {However, we numerically show that the global provider's revenue could decrease, as he shares his revenue with more providers and can only charge a lower price.} 
\end{abstract}

}

\author{Lingjie Duan, \emph{Member}, \emph{IEEE}, Jianwei Huang, \emph{Senior Member}, \emph{IEEE}, and Biying Shou \thanks{Lingjie Duan is with Engineering Systems and Design Pillar, Singapore University of Technology and Design, Singapore. Jianwei Huang is with the Network Communications and Economics Lab, Department of Information Engineering, The Chinese University of Hong Kong, Hong Kong. Biying Shou is with  the Department of Management Sciences, City University of Hong Kong, Hong Kong. Email: lingjie\_duan@sutd.edu.sg, jwhuang@ie.cuhk.edu.hk, and biying.shou@cityu.edu.hk.}
\thanks{Part of the results appeared in \emph{IEEE INFOCOM 2013} \cite{DuanHuangShouINFOCOM13}. {This work is supported by the SUTD-MIT International Design Center Grant (Project no.: IDSF1200106OH), SUTD Start-up Research Grant (Project no.: SRG ESD 2012 042), and the General Research Funds (Project Number CUHK 412710 and CUHK 412511) established under the University Grant Committee of the Hong Kong Special Administrative Region, China. This work is also partially supported by the Hong Kong General
Research Fund (Project No. CityU 144209) and grants from City
University of Hong Kong (Project No. 7008143 and 7008116).}
}
}
\maketitle

\section{Introduction}

{The 802.11 standard  based wireless local area network technology, also known as WiFi,}
is one of the most successful stories in modern wireless communications \cite{WiFi}. Operating in the unlicensed 2.4GHz {and 5GHz} spectrum band, WiFi networks do not require {exclusive}  spectrum licenses as their cellular counterparts, and can provide high-speed wireless access to mobile users within tens to hundreds of meters of WiFi access points (APs) \cite{Lehr}. Furthermore, APs in WiFi networks are inexpensive and can be easily deployed and maintained \cite{Manshaei}.
These explain why the annual revenue in the WiFi industry is growing rapidly in recent years and is expected to worth \$93.23 billion by 2018 (e.g., \!\cite{Global, Loredo}). 

In order to provide close to seamless high performance mobile communication experiences,
many WiFi providers (e.g., AT\&T in US, BT Openzone in UK, iPass in some EU countries, and PCCW in Hong Kong) are deploying a large number of WiFi APs in their local markets. {For example, iPass has set up more than 1.2 million public WiFi venues, and his revenue keeps growing 14 percent quarter-over-quarter (reaching \$20.3 milllion in the second quarter of 2013)\cite{iPass}.}  {In the Hong Kong market alone,} PCCW has increasingly rolled out more than twelve thousand public APs covering almost all popular places (e.g., convenient stores and shopping malls, coffee shops and hotels, train stations, and education institutes). {Note that some of these providers (e.g., AT\&T and PCCW) are also cellular operators; however, they provide the WiFi services separately from their cellular data plans to cater to mobile devices without intrinsic cellular connectivity (e.g., tablets and laptops) as well as users who are not their current cellular subscribers (but are willing to use their WiFi services). }{Generally, cellular data services and WiFi services target at different users: one supporting high user mobility and the other supporting high data throughput. }
{For many local providers, we often observe them charging \emph{local users (subscribers)}}
a monthly flat fee (e.g., \cite{PCCW,Orange,ATT}), where a user pays a fixed amount per month independent of the actual usage.
This motivates us to ask the first key question in this paper: \emph{Why does a local WiFi provider prefer to charge his local users a flat fee instead of a usage-based fee?}

Notice that a WiFi AP can serve not only local users, but also \emph{travelers} who {visit a particular city/country} for a short period of time.
{But paying a monthly flat fee is often not a good choice for a traveler. To cater to the needs of travelers,} {Skype  has pioneered in providing a global WiFi service under the band name of \emph{Skype WiFi}, through collaborating with many local WiFi providers who own a total of more than 1 million WiFi APs worldwide} {\cite{SkypeWiFi}}. Once a user subscribes to the Skype WiFi service, he can use any of the associated WiFi AP with his Skype account, and pays according to usage with his Skype Credit (i.e.,  ``only pays for the time you are online'', as Skype puts it). Such {flexible} Skype WiFi service provides great convenience  for travelers, but also introduces competition with local WiFi providers {among local users}. {In order to promote such cooperation, Skype needs to share part of the revenue with the local WiFi providers who provide the physical WiFi APs.} Thus, the local WiFi providers will have incentives to collaborate with Skype and share their infrastructure only if they can also gain from this new service. This motivates us to ask the second key question in this paper:
%
\emph{Why does a global provider choose usage-based pricing for his global WiFi service, and how should he share the benefits with the local WiFi providers?}

{To answer the first key question,}
we focus on a local market with a {monopolistic} local WiFi provider and a group of local users. We model their interactions as a two-stage Stackelberg game: the local provider (leader) determines pricing scheme (flat-fee or usage-based) in Stage I, and local users (followers) decide whether they will subscribe to the service (and how much to use) in Stage II.
{We show that the flat-fee pricing can offer a higher revenue than the usage-based pricing for such a monopolistic local provider.}
{To answer the second key question,} we study how the global provider may provide a global WiFi service by cooperating with local providers, {given the local providers' optimal flat-fee based pricing}. We formulate the problem as a two-stage dynamic game. In Stage I, the global provider negotiates with each local provider about the global WiFi price and the revenue sharing portion based on Nash bargaining. In Stage II,  local users choose between the global and local providers' services, and travelers choose their usage levels in the global WiFi service.

Our key results and contributions are as follows:
\begin{itemize}

\item \emph{{Flat-fee pricing dominates the local WiFi markets}}:
In Sections~\ref{subsec:LocalUsage}, \ref{subsec:LocalFlat} and \ref{Sec:local compare}, {we study the price choices of  a monopolistic local provider.} 
{We {analytically} show that the flat-rate pricing
is effective in attracting the high-valuation users, while the usage-based pricing 
is attractive to the low-valuation ones.} 
When the WiFi capacity is abundant,
{the local provider will choose the flat-fee pricing as it brings more revenue.}  

\item \emph{Win-win situation when the global provider chooses the usage-based pricing:} In Section~\ref{sec:Global}, {we {analytically} show that the global WiFi provider prefers the usage-based pricing, in order to avoid severe competition with local providers. Such pricing scheme also attracts those not served by local providers (e.g., local users with low-valuations {and travelers} {from other markets}), and hence increases the total revenue in the market.} When the revenue is shared properly, the global provider and each local provider {achieve a win-win situation}.

\item \emph{Nash bargaining on the global WiFi price and revenue sharing:} In Section \ref{subsec:SkypeUsage}, we decompose the interactions among different local markets and study each of them separately. We {analytically} show that the global provider always needs to share the majority of his revenue with local providers, {to compensate the providers' revenue loss due to competitions and incentivize them to share the infrastructure.} {If the local user population decreases or the traveler population {from other markets} increases, the global provider has a larger bargaining power and gives away less revenue .}

\item \emph{Impact of local market competition:} In Section~\ref{sec:Comp}, we extend the analysis in the monopolistic local market to a competitive market. We {analytically} show that the local provider competition reduces the market price and attracts more users. The competition provides more incentives for local providers to collaborate with the global provider, {and enables the global provider to share less revenue with each provider. However, we {numerically} show that the global provider's revenue could decrease, as he can only charge a lower price and will share revenue with more providers.}
\end{itemize}

\vspace{-5pt}
\subsection{Related Work}


%
{The recent literature on WiFi pricing can be divided into three categories.} {The first category focuses on how a local provider optimizes the price or multiple providers compete on their prices to maximize individual revenues (e.g., \cite{gizelis2011survey,niyato2008competitive, Niyato3}).} {These results often ignored the WiFi's limited coverage and the users' movements across different WiFi markets. Moreover, they often assumed either flat-rate pricing or usage-based pricing, without an analytical comparison between the two schemes.\footnote{Although Lee et al. \cite{jeonghoon2012} considered various pricing schemes, the proposed usage-based pricing does not apply to our WiFi services.}}
{The second category focused on the perspective of an individual WiFi AP owner,
who charges visitors for using his AP's resources} (e.g., \cite{musacchio2006wifi,feamster2007lease,friedman2003pricing}). {The key design challenge here is the asymmetric information, i.e.,}  visitors know more about their own utility functions than the AP owner. {{The third category {studied} wireless social community networks, where WiFi owners form a community {to share} their APs with each other, so that one AP owner can use other APs to access the Internet during travel (e.g., \cite{Manshaei,ai2009wi}).} {In {this line of literature}, the main design objective is to encourage as many AP owners to join the community as possible. The revenue maximization becomes a secondary concern.}}

{In this paper, we study the optimal pricing schemes in both local and global WiFi markets. We consider several key and practical features of WiFi networks (e.g., {WiFi's limited coverage and users' movement across different WiFi markets}), and compare the pros and cons of the flat-rate and the usage-based pricing.
Furthermore, we are the first to study
how a global WiFi service provider (such as Skype) may cooperate with local providers, negotiate pricing and revenue sharing schemes, and achieve a win-win situation. }

There are some other works studying how a monopoly provider uses a supplementary network technology to improve the existing one (e.g., using WiFi networks to offload heavy data traffic from cellular networks to avoid congestion) (e.g., \cite{Joe-Wong, Yaiparoj}). Unlike those studies, our study focuses on the public WiFi service market, and tries to understand the issue of service pricing and collaboration/competitoin locally and globally.

\vspace{-5pt}
\subsection{Taxonomy}

{The following terms will be used  throughout this paper.}

%
\begin{itemize}
\item \emph{Local provider:} {A WiFi provider who deploys APs to provide service to a single region. For example,  PCCW serves the Hong Kong market only, and AT\&T serves the USA market only.}

\item \emph{Global provider}: {A WiFi provider who serves multiple local markets, by using the network infrastructure (APs) of the corresponding local providers. For example, the Skype WiFi service  covers many countries with collaborations with local provides, but Skype does not own any physical WiFi APs.}

\item \emph{Local market:} {A market that is served by {one or multiple} local providers (and possibly by a global provider). There are a set $\mathcal{I}=\{1,2, ..., I\}$ of disjoint local markets.} {Initially we will assume that each local market has a single local provider. In Section~\ref{sec:Comp}, we will further look at the case where there are multiple local providers in the same market.} 


\item \emph{Local user:} A user who lives in a particular local market as a long-term resident. There are $N_i$ local users in each local market $i\in\mathcal{I}$.
\item \emph{Traveler:} When a user travels to a market other than his own local market, he becomes a traveler. We use the parameter {$\alpha_j^i\in[0,1]$} to denote the percentage of users in a local market $j$ who are willing  to pay short-term visits to local market $i$, and thus the total travelers from market $j$ to $i$ is $\alpha_j^iN_j$.

\end{itemize}

\section{Usage-based Pricing for Local WiFi}\label{subsec:LocalUsage}


{We will first study how a local provider in a local market $i$ optimizes the price to maximize the revenue, assuming that he chooses the usage-based pricing.}
In Section~\ref{subsec:LocalFlat}, we will {derive the optimal pricing term should the local provider choose to use the flat-fee pricing}. In Section~\ref{Sec:local compare}, we will {compare these two cases, and} show that flat-fee pricing always brings more revenue than the usage-based pricing in the local WiFi service.


We consider a two-stage {dynamic} game between a local provider $i$ and {a group of $N_i$ local users}. In Stage I, the provider $i$ announces the price $p_i$  (per unit {of usage time}) to maximize his revenue. In Stage II, users decide whether and how much to use the service to maximize their payoffs. {As there are two stages in this game and the provider is the only leader (followed by users), this is also a Stackelberg game.}
At a Subgame Perfect Equilibrium (SPE, or simply \emph{equilibrium}) of the game, the provider and users will not have incentives to change their pricing and usage choices.

Next we will analyze the {equilibrium} of the game using the backward induction \cite{fudenberg1991game}. We will first study the users' decisions in Stage II {for any given} price, and then look at how the provider should optimize the price in Stage I by taking the users' decisions into consideration.

\vspace{-10pt}
\subsection{Stage II: Users' Usage Choices}

Due to the limited number of APs, a local WiFi provider typically cannot {provide a complete coverage in a region.} Let us denote the local provider's WiFi's coverage as $G_i(M_i)\in(0,1)$, {where $M_{i}$ is the total number of deployed APs.}
In this paper, we will assume that $M_{i}$ is fixed, and thus will simply write $G_i(M_i)$ as $G_i$.
{{Notice that today's WiFi technologies support high data throughput and the coming WiFi technology IEEE 802.11ac further offers a much larger throughput (up to 866.7 Mbit/s) per user  \cite{gravogl2011choosing}. Hence, the network congestion is usually not a major issue in such WiFi networks. Furthermore, the FCC has decided to dramatically expand the unlicensed spectrum for use by WiFi devices and hence will effectively mitigate possible {WiFi congestion in the near future} \cite{FCCWiFi}.}} {Note that the WiFi deployment cost is fixed and is related to $G_i$, and the optimal pricing decisions are not affected by the cost, as long as the maximum revenue can compensate the cost.}

\del{(by Jianwei) Notation issue: when I first read the notation $d^{i}$, I thought that all users in the same local market $i$ will have the same usage level. But it is not true. Moreover, the index $i$ is a superscript in $d^{i}$ but a subscript in $p_{i}$, and this will cause confusion. Can we simply change $d^{i}$ into $d$ (and $d^{i\ast}$ into $d^{\ast}$), if such the optimal usage level is completely determined by $\theta$ and $p_{i}$ (assuming that $k$ is the same for all users in all markets)? This will not cause confusion even if we talk about multiple local markets together in later sections? (Lingjie: Done)}

When a local user  in market $i$ {is in the WiFi coverage},
we denote his usage level as $d\in [0, 1]$, which represents the percentage of Internet connection time over the whole time in WiFi coverage.\footnote{We assume that mobile users' time-varying locations follow Poisson point process (PPP), and thus each user has the same expected total time (normlized by $G_i$) within the WiFi coverage during a period of time (e.g., one month). Each user's total WiFi actual connected time is hence $G_i$ if the demand level $d=1$.} 
{For example, $d=1$ means that the user always stays online whenever  WiFi is available.}
{Different users may demand different usage levels as they have different valuations towards Internet connection.} {We characterize such a valuation}
by a type parameter $\theta$. 
{Unlike $d$, the parameter $\theta$ is not a decision variable.} 
A larger $\theta$ implies the user's higher valuation of the Internet access time. {Like many other studies in this field, we assume that $\theta$ follows a uniform distribution in $[0,1]$ for analysis tractability and the relaxation to more general distributions is unlikely to change the main engineering insights (e.g., \cite{{jeonghoon2012}, Duan-uniform, {musacchio2006wifi}}).} We further assume that a type-$\theta$ user's  utility $u(\theta,d)$ is linearly increasing in $\theta$ and concavely increasing in $d$. The {concavity assumption} is to represent his diminishing return in Internet access time. One commonly used utility function satisfying our requirement is\footnote{{The logarithmic utility is widely used in the networking literature to model elastic applications (\emph{e.g.,}\cite{courcoubetis2003pricing,duan2011investment}).}}
\begin{equation}
u(\theta,d)=\theta\ln(1+kd),
\end{equation}
where the parameter $k>0$
represents the elasticity of demand, i.e., {the ratio between} the percent change of
demand and the percent change of price \cite{mas1995microeconomic}. {In economics and marketing, the usual way to obtain the value of $k$ is through extensive market survey and statistical analysis \cite{Bell}. As it is difficult and costly to track each user's demand elasticity, it is common to examine users' aggregate behavior and use an identical $k$ for all users to represent the average elasticity. Unlike $k$, it is relatively easy to estimate the distribution of willingness to pay (i.e., $\theta$) in marketing.}

When using the service, a user needs to pay linearly proportionally to his {usage time} and the unit price $p_{i}$. {This is motivated by the fact that many providers charge based on connection time instead of data volume.} As the user's usage and payment are only meaningful when he is within the WiFi coverage, his overall payoff $v^i$ is linear in the coverage $G_{i}$,\footnote{{A mobile user will start to consider his network usage level $d$ after detecting the WiFi signal (i.e., inside the coverage of $G_i$) from time to time, and will not decide a total usage level $G_id$ beforehand. Thus we model the user's utility as $G_i\theta\ln(1+kd)$ in (\ref{eq:valuation_basis}), where the linear term $G_i$ can be view viewed as the time frequency to use $d$.}}
\begin{equation}\label{eq:valuation_basis}
v^i(\theta,p_i,d)=G_i(\theta\ln(1+kd)-p_id).
\end{equation}
Maximizing payoff $v^{i}$ over $d$ leads to the optimal usage level
\begin{equation}\label{eq:usage_demand}
d^{*}(\theta,p_i)=\min\left(\max\left(\frac{\theta}{p_i}-\frac{1}{k},0\right),1\right),
\end{equation}
which is increasing in the user's {indivisual} type $\theta$ (and the {common} elasticity parameter $k$), and is decreasing in price $p_i$. Furthermore, only users with $\theta\geq {p_i}/{k}$ will have a positive usage (i.e., subscribe to the service).

Next we exploit how users' optimal usage levels change with the price $p_{i}$. 
By assuming that the two terms in the min operation in (\ref{eq:usage_demand}) are equal at $\theta=1$, we can derive the following price threshold:
\begin{equation}
p_i^{th}=\frac{k}{k+1}.
\end{equation}
{When the price $p_{i}$ is less than $p_{i}^{th}$,  some high valuation users will choose $d(\theta,p_{i})=1$. Otherwise, all users will request a usage level less than 1 (can be zero if $\theta$ is very small). We will discuss these two scenarios in Stage I.}

\vspace{-5pt}
\subsection{Stage I: The Local Provider's Pricing Choice}

\del{I just realize that we need to remove the arrow in the figure. Indeed, the line segment should start from 0 and stops at 1, as there is no value of $\theta$ large than 1. Can you fix this for all figures? This can be more easily done if the figure is drawn in latex language, so that yo can copy paste to later figures. (It is in fact not critical, and hence can be ignored at this point if lack of time.)}

\subsubsection{Low price regime: {$p_i<  {k}/({k+1})$}}

\begin{figure}[h]
\vspace{-5pt}
\centering
\includegraphics[width=0.4\textwidth]{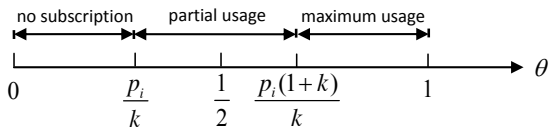}
\vspace{-5pt}
\caption{Users' WiFi usage choices in the low price regime} \label{fig:low}
\vspace{-5pt}
\end{figure}
Figure~\ref{fig:low} summarizes users' optimal usage levels in this case.
There are three categories of users based on the type parameter $\theta$: (i) a user with a small type $\theta\in[0,{p_i}/{k})$\del{do not use ``frac'' for inline maths. change later ones.} will not subscribe to the WiFi service, (ii) a user with a medium type $\theta\in\left[{p_i}/{k},{p_i(1+k)}/{k}\right)$ will subscribe with a partial usage level (i.e., $d^{*}(\theta,p_i)={\theta}/{p_i}-{1}/{k}<1$), and  (iii) a user with a high type $\theta\in\left[{p_i(1+k)}/{k},1\right]$ will have the maximum usage (\text{i.e.,} $d^{*}(\theta,p_i)=1$).
The provider's total revenue collected from the latter two user categories is
\begin{align}\label{eq:lowprice}
\pi_i(p_i)&=N_iG_ip_i\left(\int_{\frac{p_i}{k}}^{\frac{p_i(k+1)}{k}}\left(\frac{\theta}{p_i}-\frac{1}{k}\right)d\theta+\int_{\frac{p_i(k+1)}{k}}^11d\theta\right)\nonumber\\
&=N_iG_i\left(p_i-p_i^2\left(\frac{1}{2}+\frac{1}{k}\right)\right).\end{align}

By checking the first and second order derivatives of
$\pi_i(p_i)$,
we {can show that}  $\pi_i(p_i)$ is concave in $p_i$. Thus the optimal price {that maximizes the revenue in the low price regime} is
\begin{equation}\label{eq:opt_usage_price}
p_i^L=\frac{k}{k+2}.
\end{equation}
The provider's maximum revenue in the low price regime is
\begin{equation}\label{eq:opt_usage_profit}
\pi_i(p_i^L)={N_i}G_i\frac{k}{2(k+2)}.
\end{equation}

\subsubsection{High price regime: {$p_i \geq {k}/({k+1})$}}

\begin{figure}[h]
\vspace{-10pt}
\centering
\includegraphics[width=0.4\textwidth]{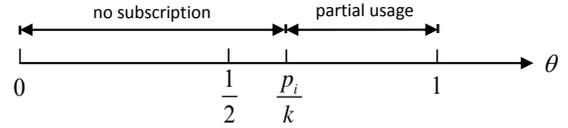}
\vspace{-5pt}
\caption{Users' WiFi usage choices in the high price regime} \label{fig:high}
\vspace{-5pt}
\end{figure}

Figure~\ref{fig:high} summarizes users' optimal usage in this case.  There are two categories of users {under such a price}: (i) a user with a low type $\theta\in[0,{p_i}/{k})$ will not subscribe to the WiFi service, and (ii) a user with a high type $\theta\in[{p_i}/{k},1]$ chooses to subscribe the WiFi service with a partial usage level (i.e., $d^{*}(\theta,p_i)={\theta}/{p_i}-{1}/{k}<1$). The provider's total revenue collected from the second user category is
\begin{align}\label{eq:usage_high}
\pi_i(p_i)&=N_iG_ip_i\int_{\frac{p_i}{k}}^1\left(\frac{\theta}{p_i}-\frac{1}{k}\right)d\theta=N_iG_i\left(\frac{1}{2}-\frac{p_i}{k}+\frac{p_i^2}{2k^2}\right).
\end{align}
The first order derivative of (\ref{eq:usage_high}) over $p_i$ is
\begin{equation}\label{eq:fakefirstorder}
\frac{d\pi_i(p_i)}{dp_i}=N_iG_i\frac{p_i-k}{k^2}.
\end{equation}
Notice that to \del{involve}obtain a positive {revenue, the provider should set the price such that the} highest type {user is} willing to subscribe, i.e., $d^{*}(1,p_i)=1/p_{i}-1/k_{i}>0$. This means $p_i<k$, which implies (\ref{eq:fakefirstorder}) is negative.  Thus the optimal price in the high price regime is $$p_i^H=\frac{k}{k+1},$$
which is the boundary case of the low price regime.

Summarizing the results from both price regimes, we have the following result.

\begin{proposition}\label{prop:local_usage}
The provider's equilibrium usage-based price {that maximizes his revenue is}
\begin{equation}\label{eq:usage_price}
p_i^*=\frac{k}{k+2},
\end{equation}
which is increasing in the elasticity parameter of demand $k$ and is independent of coverage $G_i$. The provider's maximum revenue under the equilibrium usage-based pricing is
\begin{equation}\label{eq:opt_usage_profit_final}
\pi_i(p_i^*)={N_i}G_i\frac{k}{2(k+2)}.
\end{equation}
\end{proposition}

The independence of $p_i^*$  \del{in  (\ref{eq:usage_price})} in $G_i$ is due to the fact that a user only pays when he uses the service in the WiFi coverage area.

\begin{figure}[h]
\centering
\includegraphics[width=0.4\textwidth]{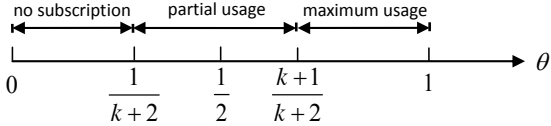}
\vspace{-5pt}
\caption{Users' WiFi usage choices at the equilibrium usage-based pricing} \label{fig:usage}
\vspace{-5pt}
\end{figure}

Figure~\ref{fig:usage} summarizes all users' usage behaviors at the equilibrium. The flexibility of usage-based pricing attracts the majority of  users to the service, since the threshold type ${p_i^*}/{k}={1}/({k+2})<{1}/{2}$. As the elasticity parameter $k$ increases, the type threshold will decrease and more users will join the service. Users' total usage level, however, is
\begin{align}\label{eq:Demand_usage}
D_i(p_i^*)&=N_iG_i\left(\int_{\frac{p_i^*}{k}}^{\frac{p_i^*(1+k)}{k}}\left(\frac{\theta}{p_i^*}-\frac{1}{k}\right)d\theta+\int_{\frac{p_i^*(1+k)}{k}}^11d\theta\right)\nonumber\\
&={N_i G_i}/{2},
\end{align}
which is independent of $k$.

\vspace{-5pt}
\section{Flat-rate Pricing for Local WiFi }\label{subsec:LocalFlat}

Similar to Section~\ref{subsec:LocalUsage}, in this section we also consider a two-stage Stackelberg game played by the provider $i$ and $N_i$ users. The difference is that the provider will announce a flat-fee in Stage I, and users decide whether to subscribe to the service in Stage II. \del{(instead of how much to use).} Since a user's payment is independent of his usage level, he will always choose {the maximum usage time} $d=1$ {(\emph{i.e.,} stay online whenever the user is in the WiFi coverage area)} whenever he subscribes. Next we derive the  game equilibrium by using the backward induction.

\vspace{-5pt}
\subsection{Stage II: Users' Subscription Choices}

In Stage II, by joining the flat-rate price plan, a type-$\theta$ user's payoff is
\begin{equation}\label{eq:user_flat}
v^i(\theta,P_i)=G_iu(\theta,1)-P_i=G_i\theta\ln(1+k)-P_i.
\end{equation}
Notice that the flat fee $P_{i}$ is independent of usage, and thus is also independent of whether the user is in the WiFi coverage area.
{In other words, once a user subscribes to the WiFi service, he will be charged a flat fee at the end of that month.} This means that the \emph{effective} price  considering the limited coverage is
$\tilde{P}_i:={P_i}/{G_i}>P_i$.

{It is clear that only} users who have high valuations of mobile Internet access would subscribe to the WiFi service {and obtain a positive payoff}. The minimum {type parameter $\theta$ among the ``active'' users} is
\begin{equation}\label{eq:partition_flat}
\theta^{th}(P_i)=\frac{P_i}{G_i\ln(1+k)}.
\end{equation}

\vspace{-5pt}
\subsection{Stage I: The Local Provider's Pricing Choice}

In Stage I, the provider wants to maximize his revenue by collecting payment from users with $\theta\in[\theta^{th}(P_i),1]$, i.e.,
\begin{equation}
\max_{P_i\geq 0}\pi_i(P_i)=N_iP_i\left(1-\frac{P_i}{G_i\ln(1+k)}\right).
\end{equation}

It is easy to verify that $\pi_i(P_i)$ is concave in $P_{i}$, and we can derive the optimal price as follows.
%

\begin{proposition}\label{prop:local_flat}
The provider's equilibrium flat-rate price that maximizes his revenue is
\begin{equation}\label{eq:price_flat}
P_i^*={G_i\ln(1+k)}/{2},
\end{equation}
which is increasing in the coverage $G_i$ and elasticity parameter $k$. The provider's maximum revenue with the equilibrium flat fee is
\begin{equation}\label{eq:revenue_Pi}
\pi_i(P_i^*)=\frac{N_i}{4}G_i\ln(1+k),
\end{equation}
which is increasing in $G_i$ and $k$.
\end{proposition}

\begin{figure}[h]
\centering
\vspace{-5pt}
\includegraphics[width=0.4\textwidth]{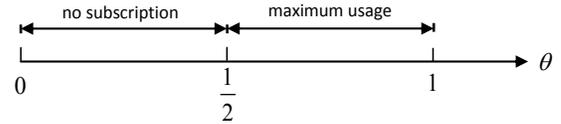}
\vspace{-5pt}
\caption{Users' WiFi usage choices at the equilibrium flat-fee pricing} \label{fig:flat}
\vspace{-5pt}
\end{figure}

Figure~\ref{fig:flat} summarizes users' usage behaviors at the equilibrium. {Comparing with Figure~\ref{fig:usage}, the inflexibility of the flat-fee pricing scheme attracts fewer users (1/2 instead of $(k+1)/(k+2)$) than the usage-based scheme.} 

Intuitively, a better WiFi coverage and a larger elasticity parameter encourage more users to join the WiFi service, and the provider can charge more. Users' total usage is
\begin{equation}
D_i(P_i^*)=\frac{N_i}{2}G_i,
\end{equation}
which is the same as in the usage-based pricing case in  (\ref{eq:Demand_usage}). {This is because users consume more (on average) with the flat-fee pricing.}

\vspace{-5pt}
\section{Flat-rate Outperforms Usage-based Pricing for Local WiFi Service}\label{Sec:local compare}

Now we are ready to {compare the two pricing schemes and see} which one leads to a larger provider revenue.

Let us define the ratio between the equilibrium revenues of the flat-rate pricing scheme and the usage-based pricing scheme as $r:={\pi_i(P_i^*)}/{\pi_i(p_i^*)}$.
Based on (\ref{eq:opt_usage_profit_final}) and (\ref{eq:revenue_Pi}), we can rewrite the ratio as a function of $k$, i.e.,
\begin{equation}
r(k)=\frac{(k+2)\ln(1+k)}{2k}.
\end{equation}

The first order derivative of $r(k)$ over $k$ is
\begin{equation}\label{eq:ratio_derivative}
\frac{dr(k)}{dk}=\frac{{k(k+2)}/{(k+1)}-2\ln(1+k)}{2k^2},
\end{equation}
and we can show that such a derivative is positive for all positive values of $k$.
%
Using L'Hospital law, we can show that
\[
\lim_{k\rightarrow 0}r(k)=\left.\frac{\ln(1+k)+{(k+2)}/{(k+1)}}{2}\right|_{k=0}=1.
\]
This means that $r(k)>1$ for any $k>0$. Thus, we have the following result.

\begin{theorem}\label{thm:flat_outperform_usage}
\del{Compared to the usage-based pricing, the provider obtains a larger revenue with the  using the flat-rate pricing.}{A local provider can obtain a larger revenue with the flat-rate pricing than with the usage-based pricing.} {The revenue gap increases in} the  elasticity parameter $k$.
\del{ increases, the revenue gap between the two pricing schemes increases.}
\end{theorem}

Theorem~\ref{thm:flat_outperform_usage} is consistent with the current industry practice, where most WiFi providers offer flat-rate pricing instead of usage-based pricing in local markets (e.g., Orange in UK \cite{Orange}, AT\&T in US \cite{ATT}, and PCCW in Hong Kong \cite{PCCW}). {Another benefit of the flat-rate pricing {(that is not explicitly modeled here)} is that it is easy to implement with little overhead for billing,}\del{Another key reason behind this is that flat-rate pricing is easy to implement with little overhead for billing,} while the usage-based pricing requires the provider to record users'  mobile traffic for payment calculation and collection over time  \cite{dasilva2000pricing}.


{\subsection{Impact of WiFi Congestion}
When a large number of users try to access the same WiFi network, they may experience network congestion, which will reduce some of their interests to join the local WiFi service. In the following, we take the congestion into account in our local WiFi model and evaluate the impact of congestion on the pricing choice. Let us denote the WiFi congestion coefficient $c(B)$, which is related to the WiFi bandwidth $B$ and models the congestion cost for one unit of WiFi demand.  We also denote the minimum type parameter $\theta$ among the WiFi subscribers as $\theta_{th}$, and users with $\theta\in[\theta_{th},1]$ will subscribe. 

\vspace{-2pt}
\subsubsection{Usage-based pricing under congestion} 
We first analyze users' best decisions in Stage II and then solve the provider's problem in Stage I by predicting users' best responses. By incorporating the congestion cost into (2), the payoff of a user with $\theta\geq \theta_{th}$ by demanding a usage level $d$ is
\begin{equation}\label{eq:payoff_congestion_usage}
v^i(\theta,p_i,d)=G_i(\theta\ln(1+kd)-p_id-cN_i\int_{\theta_{th}}^1 d^*({\theta}',p_i) d\theta'),
\end{equation}
where $d^*(\theta',p_i)$ is the optimal demand of user type-$\theta'$. As each WiFi subscriber is infinitesimal (non-atomic) in contributing to the congestion term in (\ref{eq:payoff_congestion_usage}), his optimal demand (as long as his payoff is non-negative) is not affected by the congestion and is the same as (\ref{eq:usage_demand}). That is, $d^*(\theta,p_i)=\min(\max(\theta/p_i-1/k,0),1)$. As the user with type $\theta=\theta_{th}$ is indifferent in choosing between WiFi and not, his normalized optimal payoff by $G_i$ is zero. Thus, we can derive the unique solution $\theta_{th}$ according to the following equation:
{\small\begin{align}
&v^{i*}(\theta_{th},p_i)/G_i=\theta_{th}\ln\bigg(1+k\min\bigg(\max\bigg(\frac{\theta_{th}}{p_i}-\frac{1}{k},0\bigg),1\bigg)\bigg)\nonumber\\
&-p_i\min\bigg(\max\bigg(\frac{\theta_{th}}{p_i}-\frac{1}{k},0\bigg),1\bigg)-cN_i\int_{\theta_{th}}^1 d^*(\theta',p_i)d\theta'=0,
\end{align}}
We can see that $\theta_{th}$ here becomes larger because of congestion, and fewer users will subscribe. As $\theta_{th}$ depends on the service price $p_i$, we rewrite it as $\theta_{th}(p_i)$. By predicting this, the provider's revenue-maximization problem is
\begin{equation}\label{eq:usage_congestion_prob}
\max_{p_i} \pi({p_i})= p_i N_i G_i \int_{\theta_{th}(p_i)}^1\min\left(\max\left(\frac{\theta}{p_i}-\frac{1}{k},0\right),1\right)d\theta.
\end{equation}
We can show that Problem (\ref{eq:usage_congestion_prob}) has no closed-form solution, but can be solved efficiently and numerically through an one-dimensional exhaustive search about price $p_i$. 

\subsubsection{Flat-rate pricing under congestion} 
A WiFi subscriber does not care about his contribution to the congestion and it is still optimal for him to demand a full usage level ($d=1$). Under network congestion, a type-$\theta$ user's payoff is changed from (\ref{eq:user_flat}) to
\begin{equation}
v^i(\theta,P_i)=G_i\left(\theta \ln (1+k) - cN_i\int_{\theta_{th}}^1 1d\theta\right)-P_i,
\end{equation}
which is zero for the indifferent user with $\theta=\theta_{th}$. Then user partition threshold for subscription is
\[
\theta_{th}(P_i)=\frac{P_i/G_i+cN_i}{\ln(1+k)+cN_i},
\]
which depends on the flat-rate price $P_i$ and is larger than (\ref{eq:partition_flat}) due to congestion. In Stage I, the provider's optimization problem is
\[
\max_{P_i}\pi_i(P_i)=P_iN_i\left(1-\frac{P_i/G_i+cN_i}{\ln(1+k)+cN_i}\right),
\] 
which is a concave function in $P_i$ and its optimal solution $P_i^*={G_i\ln(1+k)}/{2}$. This is the same as (\ref{eq:price_flat}) and is independent of congestion level. The resultant revenue is
\begin{equation}
\pi_i(P_i^*)=\frac{N_iG_i\ln(1+k)}{4(1+cN_i/\ln(1+k))},
\end{equation}
which is decreasing in congestion coefficient $c$. 


Now we are ready to compare the provider's optimal revenue under the two pricing schemes. As there is no closed-form solution to Problem (\ref{eq:usage_congestion_prob}), we rely on numerical results. Figure~\ref{fig:UsageFlatComparison} shows the provider's optimal revenue ratio $\pi_i(p_i^*)/\pi_i(P_i^*)$ between usage-based and flat-rate pricing. As the congestion coefficient $c$ increases or the local user population $N_i$ increases, the network congestion in the WiFi servcie increases and flat-rate pricing (not adaptive to congestion level) will eventually lose its advantage over the usage-based pricing.

\begin{figure}[tt]
\vspace{-5pt}\centering
\includegraphics[width=77mm]{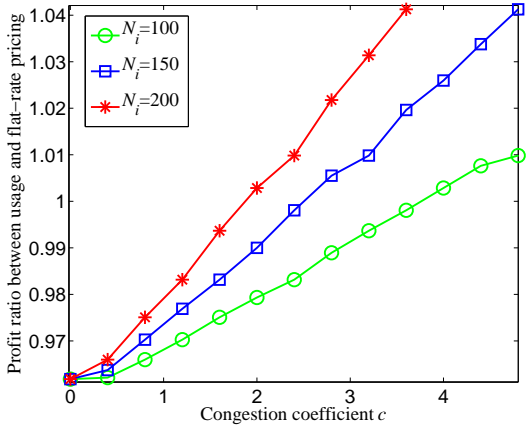}\vspace{-5pt}
\caption{The optimal profit ratio $\pi_i(p_i^*)/\pi_i(P_i^*)$ between usage-based and flat-rate pricing}\label{fig:UsageFlatComparison}
\vspace{-5pt}
\end{figure}
}

\section{Global WiFi Service}\label{sec:Global}

Now let us look at the WiFi service in a global market by investigating the fact of Skype's global WiFi operation.
If one company wants to provide a global WiFi service, he can either densely deploy APs worldwide or cooperate with many local WiFi providers. The former approach {typically} requires \del{a formidably high}{an extremely large} investment, while the latter  approach is {more} feasible. {In fact, today a global provider (e.g., Skype) uses the latter approach to provide} a global WiFi service called ``Skype WiFi'', which involves  more than 1 million APs deployed by many local providers worldwide (e.g., Best Western in US, BT Openzone in UK, and PCCW in Hong Kong). To motivate the cooperation of local providers, Skype shares some of his WiFi revenue with these cooperators \cite{Skype}. Here is what each side will gain and lose during this cooperation.
\begin{itemize}
\item \emph{Skype's gain:} Skype {can gain extra revenue by providing WiFi service.} Skype used to be just a software provider without any physical WiFi infrastructure. With the cooperation and a usage-based pricing, Skype is able to serve travelers who are not willing to sign a long-term {contract} with local providers. Furthermore, Skype can attract some low-valuation local users who do not subscribe to the flat-fee local WiFi service, or prefer usage-based pricing to the flat-fee pricing. During this process, Skype needs to share part of the revenue with the local WiFi providers to achieve a win-win situation.
\item \emph{Local provider's benefit and loss:} When Skype {starts to provide WiFi service in}\del{enters} a local market $i$, the local provider $i$ will experience {new} market competition and a reduced number of subscribers.
However, as Skype relies on the local provider's WiFi infrastructure, the provider has the market power to {negotiate} with Skype on Skype WiFi's price to avoid severe competition.\footnote{For simplicity, we assume that a local provider $i$ will still charge the same flat fee $P_i^*$ in (\ref{eq:price_flat}) after Skype's entry. In practice, a local provider may not be able to change the flat-rate price very often due to the reputation issue \cite{courcoubetis2003pricing}.\del{does my change reflect what the reference meant? Lingjie: Yes}} Furthermore, he can share part of the Skype's revenue to compensate his loss and potentially increase his total revenue.
\end{itemize}

The slogan of Skype WiFi is ``only pay for the time you're online'' (usage-based pricing). Note that Skype has the following three advantages over many local providers to implement a usage-based pricing.
\begin{itemize}
\item \emph{Existing {mechanism to record} users' traffic:} Skype can use the same traffic recording system in Skype WiFi as in the existing {Skype Internet Call Service}. 

\item \emph{Trustworthy global billing system:} Skype has built a reputable global billing platform with his existing services. As Skype has successfully cooperated with many local telecommunication companies on offering the  Skype Internet Call Service,  it is easy for Skype to convince local WiFi providers to be new collaborators.

\item \emph{High market penetration and brand visibility:} {Skype has a more than 600 million users and can  easily advertise the Skype WiFi service globally}, while many local providers have little brand visibility outside their local markets.
\end{itemize}

Even with these advantages, one may still wonder why Skype does not choose the flat-fee pricing, as what the local providers are doing for local WiFi services. Our analysis shows that one key reason is for Skype to avoid severe competition with local providers in order to reach a win-win situation.


To make the discussion more concrete, let us first look at the users' choices. After Skype's entry, a user in his own local market can choose between the local WiFi service and Skype WiFi service. When the user travels to a different market, he will only choose Skype WiFi as he does not want to pay a monthly flat fee in a different market.

Now consider the possibility of Skype adopting the flat-rate pricing scheme for the global WiFi service. This can further include two variations: a \emph{market-dependent} flat-rate pricing and a \emph{market-independent} one. In the market-dependent scheme, a user needs to pay a separate flat-rate price for each market (either local or foreign) he might enter. In this way, Skype WiFi is just replicating many local services at a global scale. This leads to direct competition with local providers in each local market (e.g., all local users in a market will choose Skype WiFi if his flat-fee is lower than the corresponding local provider's price). Furthermore, such a scheme is not attractive to a user who travels in many markets, as more markets means a higher total payment.
In the market-independent scheme, a user subscribing to Skype WiFi only needs to pay a single flat fee to receive services in all markets. Then many users no longer need to use the local WiFi service. {To summarize, in each of the two cases,} the local WiFi provider will suffer from Skype's flat-fee pricing, and will not have the incentive to cooperate. {This can explain why in practice} Skype chooses the usage-based pricing scheme.



\vspace{-5pt}
\section{Optimal Usage-based Pricing Scheme for Global WiFi provider}\label{subsec:SkypeUsage}

Now we will analyze the optimal usage-based pricing scheme for the global WiFi provider. We will consider the market-dependent usage-based pricing scheme (which is Skype's current practice), where a user pays different usage-based prices when he is in different markets.{The market-independent usage based pricing is a special case of the market-dependent one.} Under such a scheme,
%
%
we can model the {interactions between the global provider, a local provider $i$, and local users as well as travelers in market $i$}  as a two-stage dynamic game. In Stage I, the global provider and {(the local)} provider $i$ jointly decide the global WiFi usage-based price $p_i^{Glob}$ and the revenue sharing portion $\eta_i$ (as a compensation of using provider $i$'s network infrastructure). In Stage II, {each of the} $N_i$ local users {chooses between} the global provider's WiFi and the provider $i$'s local service (and the usage level if choosing global WiFi), and travelers decide their usages of the global WiFi service. {As there is more than one leader (the global and local providers) in Stage I, this game is no longer a Stackelber game but a two-stage dynamic game.} {In the following, we use backward induction to examine Stage II first.}

\vspace{-5pt}
\subsection{Stage II: Local Users' and Travelers' choices}

Consider a total of $I$ markets in the global market. A type-$\theta$ local user in the local market $i$ has two types of demands:
\begin{itemize}
\item \emph{Demand in his local market:} he can choose global WiFi's usage-based price (with the optimal usage $d^{*}(\theta,p_i^{Glob})$ as in (\ref{eq:usage_demand})) or provider $i$'s flat-rate price $P_i^*$ in (\ref{eq:price_flat}) (with the optimal maximum usage $d^{*}(\theta,P_i^*)=1$).
\item \emph{Demand when he travels in non-local markets:} he will only choose global WiFi's usage-based prices in other $I-1$ markets. The probability for this user traveling to a market $j$ is $\alpha_i^j<1$. By demanding a usage level $d^{{\ast}}(\theta,p_j^{Glob})$ in market $j$ as in (\ref{eq:usage_demand}), this user's aggregate payoff in all non-local markets is
\[
    \sum_{ j\neq i}\alpha_i^jG_j\left(u\left(\theta,d^{{\ast}}(\theta,p_j^{Glob})\right)-p_j^{Glob}d^{{\ast}}(\theta,p_j^{Glob})\right).
\]
\end{itemize}

Apparently, the user's usage in non-local markets (the second type of demand) does not affect his choice of service in the local market (the first type of demand). To study a local user's local service choice, we can simply
compare his optimal local payoff if using global WiFi, $$v^{Glob}=G_i(\theta\ln(1+kd^{*}(\theta,p_i^{Glob}))-p_i^{Glob}d^{*}(\theta,p_i^{Glob}))$$ to the optimal payoff if subscribing to the provider $i$, $$v^{i}=G_i\theta\ln(1+k)-P_i^*.$$
In the following, we analyze the local users' equilibrium behaviors given any possible value of $p_i^{Glob}$.\footnote{Note that the {revenue sharing} decision $\eta$ in Stage I does not affect users' decisions {in Stage II}.} To facilitate analysis in this section, we assume the elasticity parameter of demand $k=1$ and utility $u(\theta,d)=\theta\ln(1+d)$. {Similar to Sections~\ref{subsec:LocalUsage}, \ref{subsec:LocalFlat} and \ref{Sec:local compare}, our results here can be extended to the case with any positive $k$ value.}

\begin{figure}[tt]
\centering
\includegraphics[width=0.35\textwidth]{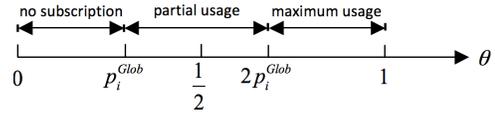}
\vspace{-5pt}\caption{{Local users' {usage}  in market $i$ with a low global WiFi price. All usage are with the global WiFi.} } \label{fig:Skype_lowprice}
\end{figure}

\begin{figure}[tt]
\vspace{-1pt}
\centering
\includegraphics[width=0.36\textwidth]{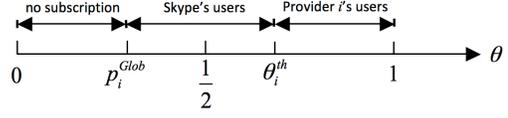}\vspace{-5pt}
\caption{Local users' service choices in market $i$ in the {medium global WiFi price regime.}} \label{fig:Skype_Provideri}
\vspace{-5pt}
\end{figure}

\begin{proposition}\label{prop:StageII}
At Stage II, local users' equilibrium decisions in market $i$ depend on the global WiFi price $p_i^{Glob}$ as follows:
\begin{itemize}
\item \emph{Low price regime ($p_i^{Glob}\leq {\ln(2)}/{2}$):}  no local users will choose provider $i$'s local service. Users with types $\theta\in[p_i^{Glob},1]$ will choose global WiFi. Their equilibrium usage levels are illustrated in Fig.~\ref{fig:Skype_lowprice}.\footnote{{The result in Fig.~~\ref{fig:Skype_lowprice} is consistent with Fig.~\ref{fig:low}, as here we set $k=1$.}}

\item \emph{Medium price regime (${\ln(2)}/{2}<p_i^{Glob}\leq {1}/{2}$):} both local provider $i$ and the global provider have positive numbers of {local} subscribers. Figure~\ref{fig:Skype_Provideri} illustrates local users' service subscriptions, where there are three categories of users depending on their valuations: (i) a user with the type $\theta\in [0, p_i^{Glob})$ will not choose any service, (ii) a user with the type $\theta\in [p_i^{Glob},\theta_i^{th})$ will choose global WiFi, and  (iii) a user with the type  $\theta\in[\theta_i^{th},1]$ will choose local provider $i$'s service. The threshold type $\theta_{i}^{th}$  is the unique solution to
\begin{equation}\label{eq:theta_th}
\theta_i^{th}\ln\left(\frac{\theta_i^{th}}{p_i^{Glob}}\right)-\theta_i^{th}+p_i^{Glob}-\left(\theta_i^{th}-\frac{1}{2}\right)\ln(2)=0,
\end{equation}
which satisfies $\theta_i^{th}<2p_i^{Glob}$ (i.e., $d^{*}(\theta_i^{th},p_i^{Glob})<1$), and $\theta_i^{th}$ is decreasing in $p_i^{Glob}$. Provider $i$'s local service targets at high-valuation users, whereas global WiFi targets at low-valuation users and none of {global WiFi subscribers} request maximum usage level.

\item \emph{High price regime ($p_i^{Glob}>{1}/{2}$):} no local users will choose global WiFi. Users with types $\theta\in\left[{1}/{2},1\right]$ choose Provider $i$'s service as in Fig.~\ref{fig:flat}.
\end{itemize}
\end{proposition}

The proof of Proposition~\ref{prop:StageII} is given in Appendix~\ref{app:StageII}. {We also provide all appendices in \cite{OurReport}.} {Note the thresholds $\ln(2)/2$ and $1/2$ identify whether the global WiFi price is low enough to attract all local users or high enough to attract no local users, respectively. Both thresholds are less than $1$ as they cannot exceed the maximum user type $\theta=1$.}
Proposition~\ref{prop:StageII} shows that
two services will coexist only in the medium {global WiFi} price regime, when the two prices are comparable to each other.
When $p_i^{Glob}$ decreases in this regime,  more local users will switch from the local provider $i$ to global WiFi, resulting in a larger partition threshold $\theta_i^{th}$. Moreover, $\theta_i^{th}$ only depends on $p_i^{Glob}$ and is independent of WiFi coverage $G_i$, as both services compete with each other using the same network infrastructure.

\vspace{-5pt}
\subsection{Stage I: Negotiation Between Global and Local Providers}\label{subsec:bargain}

As the low and high price regimes in Proposition~\ref{prop:StageII} will drive either local provider $i$ or the global provider out of the local market, they are not likely to be viable choices for the negotiation in Stage I. In fact, we can prove that the medium price regime is the only {practical} choice for the whole game equilibrium.
\begin{theorem}\label{thm:win_win}
In Stage I, the global provider and local provider $i$ will only agree on a global WiFi price in the medium price regime (i.e., ${\ln(2)}/{2}<p_i^{Glob}\leq {1}/{2}$) {as long as the local user number is nontrivial compared to the traveler number from other markets}.\footnote{{We rule out the extreme case where the local user number is trivial compared to the traveler number (i.e., ${N_{i}}/{\left(\sum_{j\neq i}\alpha_{j}^{i}N_{j}\right)}\rightarrow 0$), in which case the global provider will become the monopolist in market $i$ with the monopoly price of $1/3$ (in the low price regime) and serve the travelers only. It is clear that in the majority of  markets the number of local users should be much larger (at least comparable) to the traveler number. Actually, a small number of local users cannot compensate the initial deployment cost of a large-scale WiFi network and does not allow the existence of a local provider in the first place (before the global provider's entry). However, for the purpose of completeness, we still provide the analysis for this extreme case in Appendix~\ref{app:lowprice}.}}
\end{theorem}

{The proof Theorem~\ref{thm:win_win} is given in Appendix~\ref{app:lowprice}.}
%
Next we focus on the medium price regime and study the revenues for both the global provider and Provider $i$.

\begin{figure}[tt]
\centering
\includegraphics[width=0.35\textwidth]{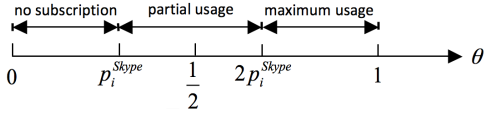}
\vspace{-5pt}
\caption{Travelers' usage choices in market $i$ in global WiFi in the medium price regime} \label{fig:Skype_mediumprice_travelers}
\vspace{-10pt}
\end{figure}


\begin{table*}[tt]
\begin{multline}\label{eq:eta_pi}
\eta_i^*(p_i^{Glob})=\min\left(1,\frac{\frac{\ln(2)}{4}N_i\left(\theta_i^{th}(p_i^{Glob})-\frac{1}{2}\right)}{\sum_{ j\neq i}\alpha_j^iN_j\left(p_i^{Glob}-\frac{3(p_i^{Glob})^2}{2}\right)+N_i\frac{(\theta_i^{th}(p_i^{Glob})-p_i^{Glob})^2}{2}}\right.\\
+\left.\max\left(\frac{\frac{\ln(2)}{4}N_i\left(\theta_i^{th}(p_i^{Glob})-\frac{1}{2}\right)}{\sum_{ j\neq i}\alpha_j^iN_j\left(p_i^{Glob}-\frac{3(p_i^{Glob})^2}{2}\right)+N_i\frac{(\theta_i^{th}(p_i^{Glob})-p_i^{Glob})^2}{2}},\frac{1}{2}\right)\right),
\end{multline}
\hrule 
\vspace{-7pt}
\end{table*}

First consider the global provider, who gains revenue by serving both local users and travelers in market $i$, but needs to share $\eta_i$ portion {of his  revenue from market $i$} to local provider $i$ for using the local WiFi infrastructure. {Global WiFi's revenue from other markets is not related to local provider $i$, and can be  normalized to 0 in the following analysis.}
By serving local users with types $\theta\in[p_i^{Glob},\theta_i^{th})$, the global provider collects a total payment
\begin{multline}\label{eq:DeltaSkype_i}\Delta \pi_i^{Glob}(\eta_i,p_i^{Glob},\theta_i^{th})\\=(1-\eta_i)p_i^{Glob}N_iG_i\int_{p_i^{Glob}}^{\theta_i^{th}}d^{*}(\theta,p_i^{Glob})d\theta.
\end{multline}
with  $d^{*}(\theta,p_i^{Glob})={\theta}/{p_i^{Glob}}-1$.
%
As for travelers in market $i$, they can be divided into  three categories depending on their valuations (and independent of where they come from) as in Fig.~\ref{fig:Skype_mediumprice_travelers}: \del{Wrong figure? need a new figure?} (i) travelers  with types $\theta\in[0,p_i^{Glob})$ demand zero usage, (ii) travelers  with types $\theta\in[p_i^{Glob},2p_i^{Glob})$ demand partial usage, and (iii)  travelers with types  $\theta\in[2p_i^{Glob},1]$ demand the maximum usage. Similar to (\ref{eq:lowprice}), we can derive that the total payment collected by the global provider from the travelers in market $i$ as
\begin{align}\label{eq:DeltaSkype_-i}
\Delta \pi_{-i}^{Glob}(\eta_i,p_i^{Glob},\theta_i^{th})=&(1-\eta_i)\sum_{ j\neq i}\alpha_j^iN_jG_ip_i^{Glob}\nonumber\\
&\! \! \! \! \! \! \! \! \! \! \! \! \! \! \! \! \! \! \! \! \! \! \! \! \! \! \! \! \! \! \! \! \! \!  \! \! \! \! \! \!  \! \! \! \! \! \!  \! \! \! \! \! \! \! \! \! \! \! \! \! \!\!\!\! \cdot\left(\int_{p_i^{Glob}}^{2p_i^{Glob}}\left(\frac{\theta}{p_i^{Glob}}-1\right)d\theta+\int_{2p_i^{Glob}}^11d\theta\right).
\end{align}
By summing up (\ref{eq:DeltaSkype_i}) and (\ref{eq:DeltaSkype_-i}), the global provider's revenue increase ({comparing with the zero revenue if he does not cooperate}) by cooperating with local provider $i$ is
\begin{align}\label{eq:SkypeBenefit}
\Delta\pi^{Glob}(\eta_i,p_i^{Glob},\theta_i^{th})=&(1-\eta_i)N_iG_i\frac{(\theta_i^{th}-p_i^{Glob})^2}{2}\nonumber\\
&\! \! \! \! \! \! \! \! \! \! \! \! \! \! \! \! \! \! \! \! \! \! \! \! \! \! \! \! \! \! \! \! \! \!  \! \! \! \! \! \!  \! \! \! \! \! \!  \! \! \! \! \! \! \!\!\!\!\!\!\!\!\!+(1-\eta_i)\sum_{ j\neq i}\alpha_j^iN_jG_i\left(p_i^{Glob}-\frac{3(p_i^{Glob})^2}{2}\right),
\end{align}
which linearly decreases in the revenue sharing portion $\eta_i$ and increases in the number of travelers $\sum_{j\neq i}\alpha_j^iN_j$ from the other markets.\del{(by Jianwei) This following sentence is not clear. I though the decoupling was explained in footnote \ref{footnote:decomposition}. Since this is an important point that we mentioned in both introduction and abstract, we need to explain it more carefully.} {Notice that the number of travelers  in market $i$ is fixed, and (\ref{eq:SkypeBenefit}) is independent of other local markets' operations.\footnote{{At the equilibrium, each local provider will join the collaboration with the global provider and realize a win-win situation. Thus we can study each market individually by presuming the global provider's collaborations with all other providers.}} Thus we can decompose the interactions between different local markets, and study each of them separately.} {Note that a local provider's revenue and the global provider's local revenue are still dependent on the number of travelers from other markets. }

\begin{figure*}[tt]
   \begin{minipage}[t]{0.32\linewidth}\vspace{-5pt}
      \centering
      \includegraphics[width=1.08\textwidth]{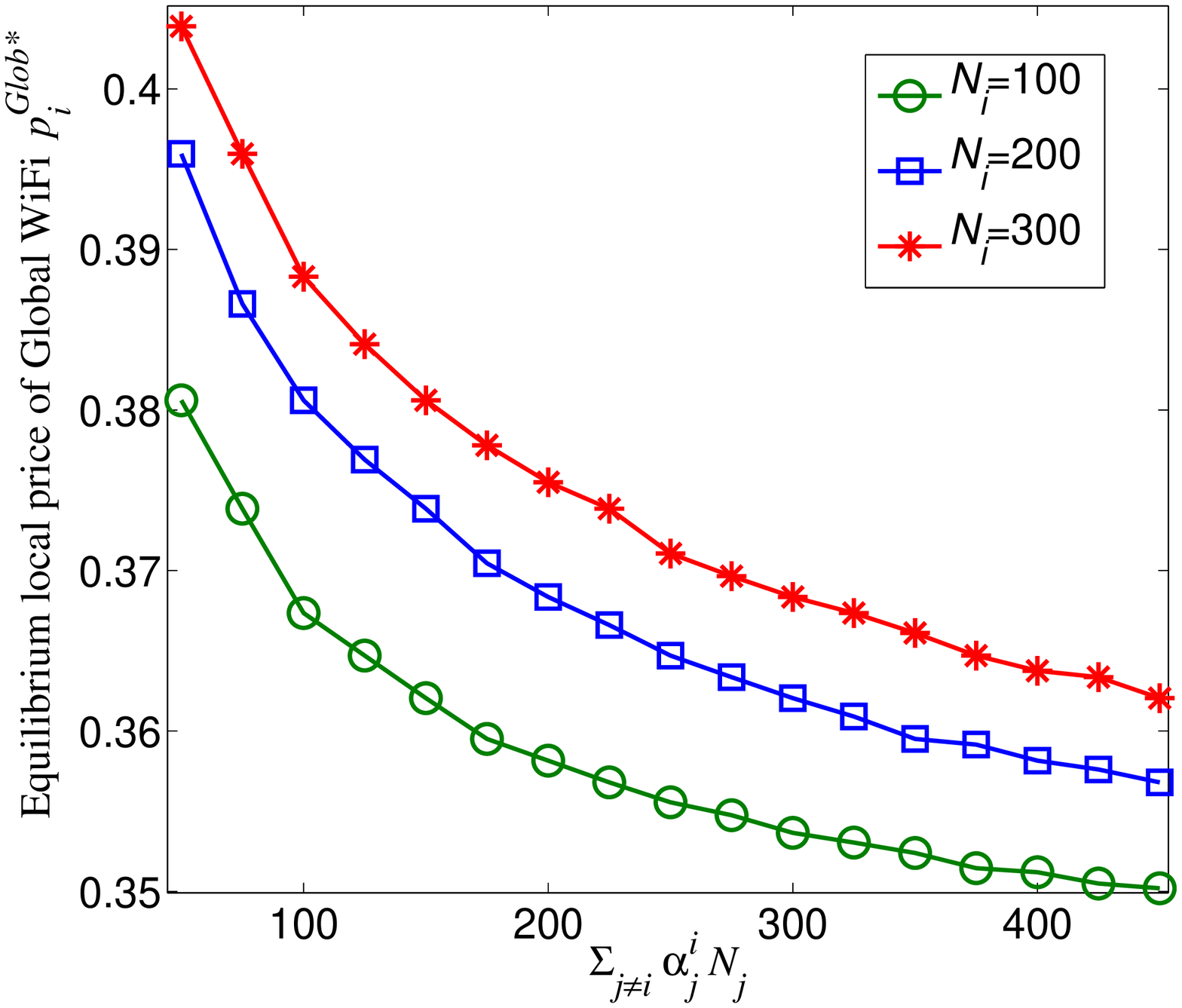}\vspace{-5pt}
      \caption{Equilibrium price of global WiFi $p_i^{Glob*}$ in market $i$} \label{fig:Skype_Provideri_price}\vspace{-5pt}
   \end{minipage}
   \begin{minipage}[t]{0.32\linewidth}\vspace{-5pt}
      \centering
      \includegraphics[width=1.1\textwidth]{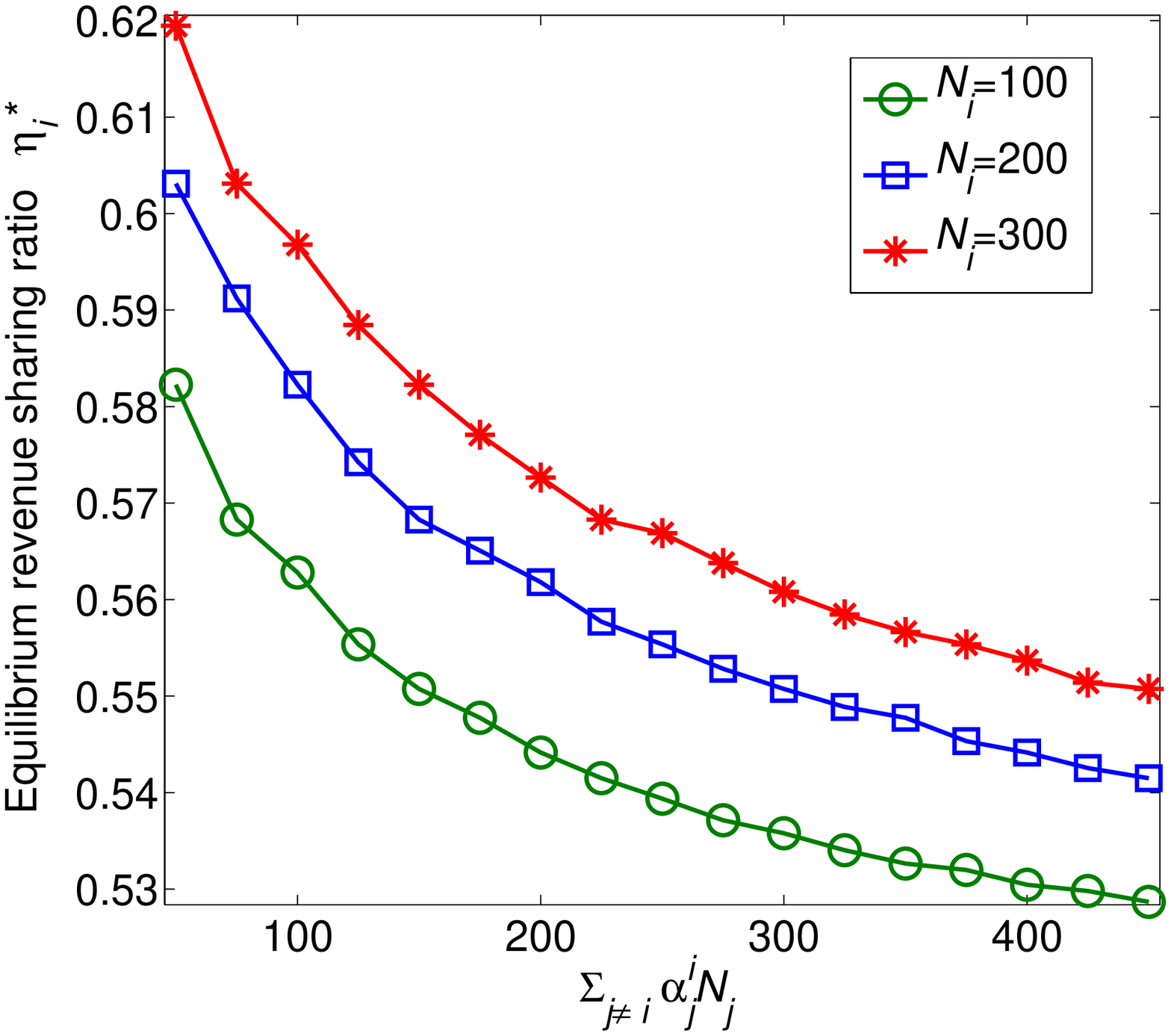}\vspace{-5pt}
      \caption{Sharing portion $\eta_i^*$ between Provider $i$ and the global provider} \label{fig:Skype_Provideri_ratio}\vspace{-5pt}
   \end{minipage}
   \begin{minipage}[t]{0.32\linewidth}\vspace{-5pt}
      \centering
      \includegraphics[width=1.1\textwidth]{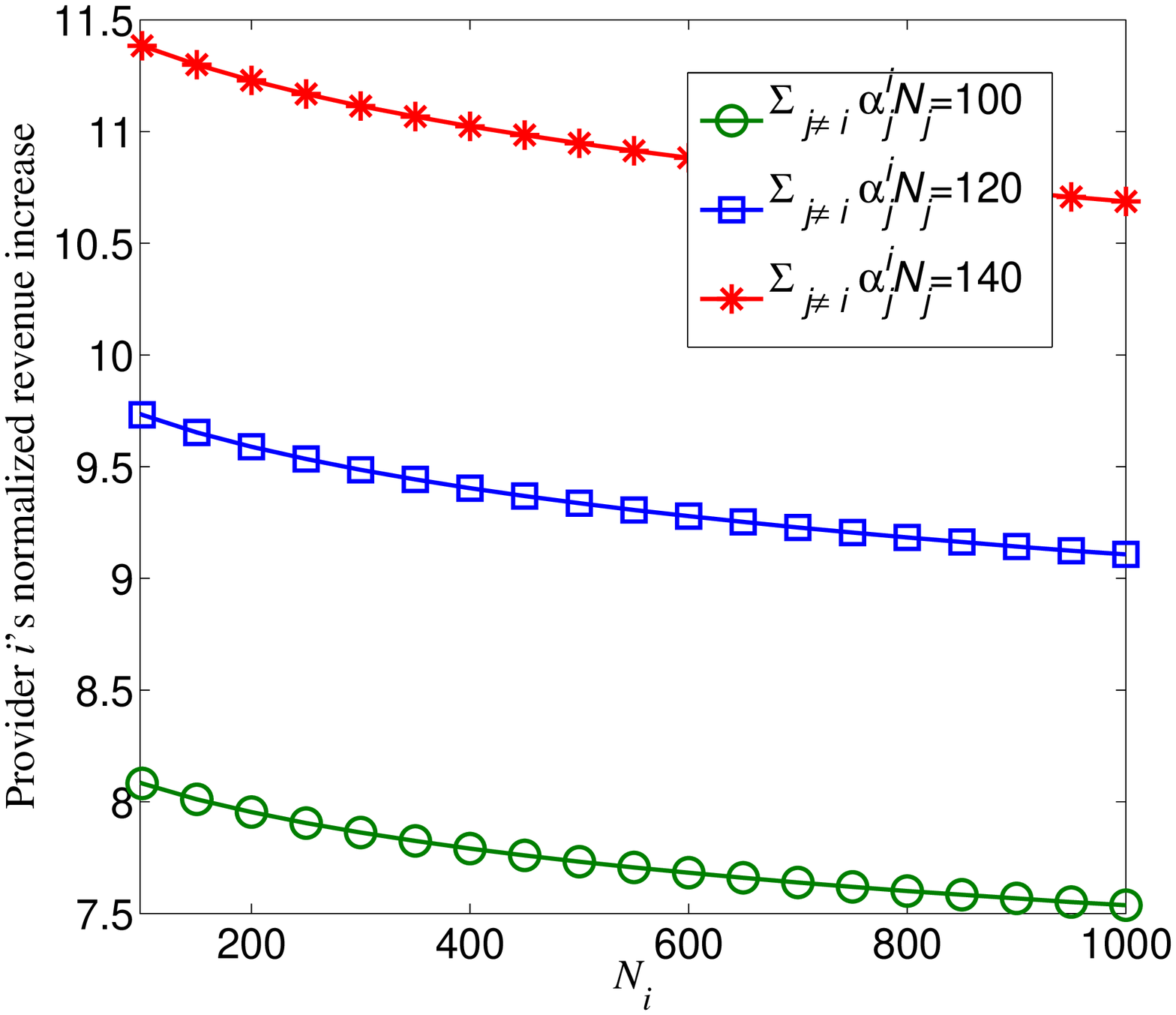}\vspace{-5pt}
\caption{Provider $i$'s normalized equilibrium revenue increase by $G_i$.}
\label{fig:Provider_i_Ni}\vspace{-5pt}
   \end{minipage}
\end{figure*}

Now we look at the revenue increase of local provider $i$ through the cooperation. As the global provider's entry will result in service competition, provider $i$ will lose those users with types $\theta\in[{1}/{2},\theta_i^{th}]$ to global WiFi. Compared with provider $i$'s original revenue in (\ref{eq:revenue_Pi}) with $k=1$, such competition reduces the  revenue by
\begin{equation}\Delta\pi_i^i(\eta_i,p_i^{Glob},\theta_i^{th})=-\frac{\ln(2)}{2}G_iN_i\left(\theta_i^{th}-\frac{1}{2}\right)<0.\nonumber\end{equation}
On the other hand, the global provider will share part of his revenue with local provider $i$:
%
$$\Delta\pi_{-i}^{i}(\eta_i,p_i^{Glob},\theta_i^{th})=\frac{\eta_i}{1-\eta_i}\Delta\pi^{Glob}(\eta_i,p_i^{Glob},\theta_i^{th}).$$
Thus local provider $i$'s total revenue increase is
\begin{align}\label{eq:Provider_i_Benefit}
&\Delta\pi^{i}(\eta_i,p_i^{Glob},\theta_i^{th})\nonumber \\
& = \Delta\pi_i^i(\eta_i,p_i^{Glob},\theta_i^{th}) + \Delta\pi_{-i}^{i}(\eta_i,p_i^{Glob},\theta_i^{th}) \nonumber \\
&=-\frac{\ln(2)}{2}G_iN_i\left(\theta_i^{th}-\frac{1}{2}\right)+\eta_iG_i\nonumber\\
&\times \left(N_i\frac{(\theta_i^{th}-p_i^{Glob})^2}{2}+\sum_{ j\neq i}\alpha_j^iN_j\left(p_i^{Glob}-\frac{3(p_i^{Glob})^2}{2}\right)\right),
\end{align}
which linearly increases in $\eta_i$ and $\sum_{ j\neq i}\alpha_j^iN_j$.

Now we discuss how the global provider bargains with local provider $i$ on $p_i^{Glob}$ and $\eta_i$ based on (\ref{eq:SkypeBenefit}) and (\ref{eq:Provider_i_Benefit}).
We will use the Nash bargaining framework to resolve this issue. According to \cite{fudenberg1991game}, the Nash bargaining equilibrium is Pareto efficient, {symmetric, and independent of irrelevant alternatives}.\del{add several other axioms of NBS here.} It is the same as Zeuthen's solution of a general bargaining problem where two players could bargain for infinite rounds. In our problem, the Nash bargaining leads to the following joint optimization problem of the revenue increase product,\footnote{We can add different weights to each term in the product to reflect different market powers of the global provider and provider $i$, but this will not change the key insights of this paper.}
\begin{align}\label{eq:prob_Skype_Provider_i}
\max_{\eta_i,p_i^{Glob},\theta_i^{th}}&\ \Delta \pi^{Glob}(\eta_i,p_i^{Glob},\theta_i^{th})\Delta\pi^{i}(\eta_i,p_i^{Glob},\theta_i^{th})\nonumber\\
\text{subject\ to,\ } &\ \ \ \ \ \ \ \ \ \quad \quad 0\leq \eta_i\leq 1, \nonumber\\
&\ \ \ \ \ \ \quad \frac{\ln(2)}{2}\leq p_i^{Glob}\leq \frac{1}{2},\nonumber\\
&\! \! \! \! \! \! \! \! \! \! \! \! \! \! \! \! \! \! \! \! \! \!\! \theta_i^{th}\ln\left(\frac{\theta_i^{th}}{p_i^{Glob}}\right)-\theta_i^{th}+p_i^{Glob}-\left(\theta_i^{th}-\frac{1}{2}\right)\ln(2)=0,
\end{align}
where the last constraint comes from (\ref{eq:theta_th}). Notice that $\theta_i^{th}$ only depends on $p_i^{Glob}$, and thus we can express it as $\theta_i^{th}(p_i^{Glob})$\del{ which is included but not explicitly shown in the objective function}.
This means that we need to solve the remaining two variables $\eta_i$ and $p_i^{Glob}$ in Problem~(\ref{eq:prob_Skype_Provider_i}).
We can take a sequential optimization approach: first optimize over $\eta_i$ given a fixed $p_i^{Glob}$, and then
optimize over $p_i^{Glob}$. We can show that the objective function of Problem~(\ref{eq:prob_Skype_Provider_i}) is strictly concave in $\eta_i$, which leads to the following result.

\begin{proposition}\label{prop:eta}
At the equilibrium, the global provider shares the \emph{majority} of his revenue in market $i$ with local provider $i$, i.e., $\eta_i^*>{1}/{2}$.
 More specifically, given any feasible price ${\ln(2)}/{2}<p_i^{Glob}<{1}/{2}$, the unique optimal $\eta_i^*(p_i^{Glob})$ is given in (\ref{eq:eta_pi}),
which increases in the local user population $N_i$ and decreases in the traveler population $\sum_{ j\neq i}\alpha_j^iN_j$ {from other markets} to market $i$.
\end{proposition}

{It is interesting to observe that the global provider always needs to give away more than half of his revenue to the local provider in order to provide enough incentives for cooperation.}
%
As the local user population $N_i$ increases, {the negative impact of competition increases, hence the global provider needs to give away more revenue.}
{On the other hand,} as more travelers coming, the {relative} importance of the local market decreases, and hence the global provider can {keep more revenue} (but still less than half).

With (\ref{eq:eta_pi}), we can simplify Problem~(\ref{eq:prob_Skype_Provider_i}) into the following one variable optimization problem:
\begin{align}\label{eq:prob_Skype_Provider_i_simple}
&\max_{p_i^{Glob}}\Delta \pi^{Glob}\left(\eta_i^*(p_i^{Glob}),p_i^{Glob},\theta_i^{th}(p_i^{Glob})\right)\nonumber\\
&\quad \quad \quad \cdot \Delta\pi^{i}\left(\eta_i^*(p_i^{Glob}),p_i^{Glob},\theta_i^{th}(p_i^{Glob})\right)\nonumber\\
&\text{subject\ to,\ } \ \ \ \ \  \ \ \ \ \frac{\ln(2)}{2}\leq p_i^{Glob}\leq \frac{1}{2},
\end{align}
where $\eta_i^*(p_i^{Glob})$ is given in (\ref{eq:eta_pi}) and $\theta_i^{th}(p_i^{Glob})$ (though not in closed-form) can be derived from (\ref{eq:theta_th}). We can check that the objective of Problem~(\ref{eq:prob_Skype_Provider_i_simple}) may not be concave in $p_i^{Glob}$ and Problem~(\ref{eq:prob_Skype_Provider_i_simple}) is not a convex optimization problem. Despite this, we can still use an efficient one-dimensional exhaustive search algorithm to find the global optimal solution $p_i^{Glob*}$ \cite{ruszczynski2011nonlinear}.\footnote{{Here is an algorithm to solve Problem~(\ref{eq:prob_Skype_Provider_i_simple}). We first approximate the continuous feasible range $[\ln(2)/2, 1/2 ]$ of $p_i^{Glob}$ through a proper discretization with gap $\Delta$, i.e., representing all possibilities by $\frac{1-\ln(2)}{2\Delta}$ equally spaced values (with the first and last values equal to $\ln(2)/2$ and $1/2$, respectively). By comparing their corresponding objective values, we then determine $p_i^{Glob*}$. The overall computation complexity is $\mathcal{O}(\frac{1-\ln(2)}{2\Delta})$. In practice, the global provider will not change $p_i^{Glob*}$ frequently, and there is no need to solve Problem (\ref{eq:prob_Skype_Provider_i_simple}) often.}} {Next we highlight some key observations of the solutions to Problem (\ref{eq:prob_Skype_Provider_i_simple}).}



\begin{observation}
At the equilibrium of market $i$, both the global WiFi price $p_i^{Glob*}$ and revenue sharing portion $\eta_i^*$ 
are independent of the local WiFi coverage $G_i$.\footnote{{As both $\Delta\pi^{Glob}$ in (\ref{eq:SkypeBenefit}) and $\Delta\pi^i$ in (\ref{eq:Provider_i_Benefit}) are linear in $G_i$, and $\theta_i^{th}$ is independent of $G_i$ according to (\ref{eq:theta_th}), the objective of Problem (\ref{eq:prob_Skype_Provider_i_simple}) can be normalized over $G_i^2$. Thus the optimal solutions $p_i^{Glob*}$ and $\eta_i^*$ to the problem are also independent of $G_i$.}} As the local user population $N_i$ decreases or the traveler population $\sum_{ j\neq i}\alpha_j^iN_j$ increases, both $p_i^{Glob*}$ and $\eta_i^*$  decrease (see Figs.~\ref{fig:Skype_Provideri_price} and \ref{fig:Skype_Provideri_ratio}). 
\end{observation}

\del{The y-axis labels of Figs.~\ref{fig:Skype_Provideri_price} and ~\ref{fig:Skype_Provideri_ratio} need to be consistent with the texts: $p_i^{Glob*}$ and $\eta_i^*$.}

\del{As both provider $i$'s and the global provider's revenue increases in Problem~(\ref{eq:prob_Skype_Provider_i_simple}) are linear in coverage $G_i$, the solutions $p_i^{Glob*}$ and $\eta_i^*$ are independent of $G_i$.}\del{It is not clear why the previous sentence is obvious.} 
Note that the global provider is the monopolist for travelers, whereas both the global provider and provider $i$ are competing in serving local users. Compared to the monopoly usage-based price ${1}/{3}$ in (\ref{eq:usage_price}) with $k=1$, the price of global WiFi $p_i^{Glob*}$ needs to be higher than $1/3$ to avoid severe price competition with  provider $i$'s local flat-rate pricing service. When the traveler population $\sum_{j\neq i}\alpha_j^iN_j$ increases or local user population $N_i$ decreases, the global provider is gaining a market power approaching a monopolist in serving the whole market, and it is efficient for the global provider to lower the price {(and eventually} approach the monopoly benchmark of $1/3$ {as shown in Fig.~\ref{fig:Skype_Provideri_price}}). Meanwhile, local provider $i$'s loss of revenue due to the global provider's competition is smaller, and the global provider only needs to share a smaller portion $\eta_i^*$ with provider $i$ {as shown in Fig.~\ref{fig:Skype_Provideri_ratio}}.

\begin{observation}\label{ob:monopoly_revenue}
The equilibrium revenue increases of both provider $i$ and the global provider, {$\Delta\pi^{Glob}$ and $\Delta\pi^{i}$,} are increasing in $\sum_{j\neq i}\alpha_j^iN_j$ and $G_i$, but are decreasing in $N_i$ (see Fig.~\ref{fig:Provider_i_Ni}).
\end{observation}

\del{I remove the $\forall$ symbol in the summation $\sum_{ j\neq i}\alpha_j^iN_j$, as it should be clear from convention. You might want to adjust this in Fig.~\ref{fig:Provider_i_Ni}. }

Intuitively, a larger coverage $G_i$ improves the quality of both two services, and a larger $\sum_{j\neq i}\alpha_j^iN_j$ provides a larger  cooperation benefit between the global provider and local provider $i$. However, a larger population $N_i$ increases the competition  between the global and local providers and thus reduces the cooperation benefit.

\section{Impact of local Market competition}\label{sec:Comp}
In previous sections, we have assumed that {there is a single local provider in each local market}, as very few providers can afford the very high cost to deploy a large scale WiFi network. In this section, we relax this assumption and consider two competitive providers $1$ and $2$ in a local market $i\in\mathcal{I}$.\footnote{{As the general case of oligopoly (which involves more than two local providers in the same local market) is quite challenging to analyze, we focus on the case of  duopoly which already provides significant engineering insights for our problem.} 
} We would like to understand how local competition affects local providers' pricing strategies and  the global provider's entry into the local market.

\vspace{-5pt}
\subsection{Duopoly competition in WiFi pricing}
Let us denote the two local providers' WiFi coverages as $G_{i,1}$ and $G_{i,2}$, and we can assume $G_{i,1}\geq G_{i,2}$ without loss of generality. \del{Do we know that the two providers will definitely choose flat-fee pricing instead of usage-based pricing? Lingjie: See the following footnote 14} {If a provider $j(=1,2)$ announces a flat-fee price  $P_{i,j}$,\footnote{{Another possible scenario can be that one provider uses flat-rate pricing and the other provider uses usage-based one, as at least one provider wants to use the efficient flat-rate pricing. This scenario can be analyzed in a similar way as the local competition between the global provider and a provider in Section~\ref{subsec:SkypeUsage}, hence we skip the details here. In practice, we observe flat-rate pricing in most competition markets, as the usage-based pricing is complex and costly to manage.}} then a type-$\theta$ user's payoff by choosing provider $j$ is
\[v^{i,j}(\theta)=\theta G_{i,j}\ln(2)-P_{i,j},\  j=1,2.\]
The user will choose the provider that offers a larger payoff:
\begin{equation}\label{eq:userchoiceduopoly}
j^{\ast}(\theta)= \arg\max_{j=1,2}v^{i,j}(\theta).
\end{equation}
If  payoffs obtained from both providers are the same, the user will randomly choose one provider with a   probability of 1/2. 

Given users' preferences, two providers will optimize their prices in order to achieve an equilibrium, where each provider is maximizing his revenue given the price of the other provider. Next we will characterize the equilibrium flat-fee prices.}

We first consider the symmetric case $G_{i,1}=G_{i,2}$. {By showing that each provider wants to reduce his price to be lower than his competitor and any reasonable price should be non-negative, we have the following result.} 

\begin{proposition}\label{prop:symmetric}
Given a symmetric WiFi coverage $G_{i,1}=G_{i,2}$, {the unique equilibrium flat-rate prices are $P_{i,1}^*=P_{i,2}^*=0$.}
\end{proposition}

{Proposition \ref{prop:symmetric} is the same as the non-profitable pricing equilibrium in the classic Bertrand model of perfect competition\cite{mas1995microeconomic}.} At the equilibrium, all users will subscribe to the WiFi service, and the total demand in the market is equally shared by the two providers. A type-$\theta$ user obtains a payoff of $\theta\frac{G_i}{2}\ln(2)>0$, and each provider obtains a payoff of 0.

Next we consider the asymmetric case $G_{i,1}>G_{i,2}$. 

\begin{figure}[tt]
\vspace{-5pt}
\centering
\includegraphics[width=0.4\textwidth]{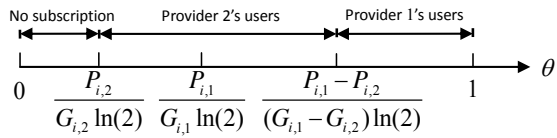}\vspace{-5pt}
\caption{Local users' choices between providers 1 and 2 in market $i$}\vspace{-5pt} \label{fig:Providerij_competition}
\vspace{-3pt}
\end{figure}

\begin{lemma}\label{lem:relationship}
{Given an asymmetric WiFi coverage $G_{i,1}>G_{i,2}$, the equilibrium prices $(P_{i,1},P_{i,2})$ satisfy}
\begin{equation}\label{eq:relation_competition}
0<\frac{P_{i,2}}{G_{i,2}\ln(2)}<\frac{P_{i,1}}{G_{i,1}\ln(2)}<\frac{P_{i,1}-P_{i,2}}{(G_{i,1}-G_{i,2})\ln(2)}<1,
\end{equation}
and the local users' subscriptions are shown in Fig.~\ref{fig:Providerij_competition}.
\end{lemma}

The proof of Lemma~\ref{lem:relationship} is given in Appendix~\ref{app:relationship}.
Using Lemma~\ref{lem:relationship}, we can derive  two providers' revenues:
\begin{equation}\label{eq:comp_rev_1}
\pi_{i,1}(P_{i,1},P_{i,2})=P_{i,1}N_i\left(1-\frac{P_{i,1}-P_{i,2}}{(G_{i,1}-G_{i,2})\ln(2)}\right),
\end{equation}
and
\begin{equation}\label{eq:comp_rev_2}
\pi_{i,2}(P_{i,1},P_{i,2})=P_{i,2}N_i\left(\frac{P_{i,1}-P_{i,2}}{(G_{i,1}-G_{i,2})\ln(2)}-\frac{P_{i,2}}{G_{i,2}\ln(2)}\right).
\end{equation}
We can show that both  $\pi_{i,1}(P_{i,1},P_{i,2})$ in (\ref{eq:comp_rev_1}) and $\pi_{i,2}(P_{i,1},P_{i,2})$ in (\ref{eq:comp_rev_2}) are {jointly} concave in $P_{i,1}$ and $P_{i,2}$. By checking the first-order conditions, we can derive each provider's {best response price (i.e., the price that maximizes his revenue given his competitor's price)}, i.e.,
\begin{equation}\label{eq:bestresponseprovider1}
P_{i,1}^*(P_{i,2})=\frac{(G_{i,1}-G_{i,2})\ln(2)}{2}+\frac{P_{i,2}}{2},
\end{equation}
and
\begin{equation}\label{eq:bestresponseprovider2}
P_{i,2}^*(P_{i,1})=\frac{G_{i,2}}{2G_{i,1}}P_{i,1}.
\end{equation}
{By solving (\ref{eq:bestresponseprovider1}) and (\ref{eq:bestresponseprovider2}) simultaneously, we obtain the unique pricing equilibrium as follows.}

\begin{figure}[tt]
\vspace{-1pt}
\centering
\includegraphics[width=0.43\textwidth]{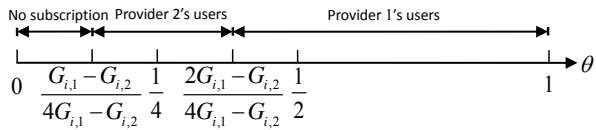}\vspace{-5pt}
\caption{Local users' equilibrium choices between providers 1 and 2 in market $i$}\vspace{-5pt} \label{fig:Providerij_competition_eq}
\vspace{-3pt}
\end{figure}

\begin{theorem}\label{thm:asymmetric}
Given an asymmetric WiFi coverage $G_{i,1}>G_{i,2}$, the unique equilibrium flat-rate prices are
\begin{equation}
P_{i,1}^*=\frac{2\ln(2)G_{i,1}(G_{i,1}-G_{i,2})}{4G_{i,1}-G_{i,2}},
\end{equation}
and
\begin{equation}
\ P_{i,2}^*=\frac{\ln(2)G_{i,2}(G_{i,1}-G_{i,2})}{4G_{i,1}-G_{i,2}}.
\end{equation}
\end{theorem}

Both equilibrium prices are lower than the monopoly price ${(G_{i,1}+G_{i,2})\ln(2)}/{2}$ in (\ref{eq:price_flat}), {if we assume that the monopolist has a coverage of $G_{i,1}+G_{i,2}$}.\footnote{One can also show that the two competitive prices are {still} lower than the monopoly price {even if} the monopolist only covers $G_{i,1}$.}
{In the monopoly equilibrium, a monopolist with the optimal flat-rate pricing can only serve 50\% of users (see Fig.~\ref{fig:flat})}; {in the duopoly case here, however,} the two competitive providers {together serve} more than 75\% of users, and provider 1 alone serves more than 50\% of users (see Fig.~\ref{fig:Providerij_competition_eq}).

Intuitively, the local market competition significantly drives the market price down and {attracts} more users. Under asymmetric service qualities {(represented by coverages)}, two providers can still differentiate their prices to cater to different groups of users and make profits. {However, under a symmetric service quality, the severe competition will bring their profits down to zero.}\footnote{{We want to remind the readers that the zero profit result can be best understood \emph{qualitatively} (\emph{i.e.}, the profits are very small), as our model does not capture all factors that may affect the providers' profits in a real market. In fact, most analytical results in this paper should be understood similarly.}}

{It should be pointed out that though we only look at duopoly case, the analysis in this subsection can be extended to oligopoly case. For example, by showing that each provider with identical coverage will reduce price to be lower than his competitors, we have the same result about non-profitable pricing as in Proposition~\ref{prop:symmetric}. Like Lemma~\ref{lem:relationship} and Theorem~\ref{thm:asymmetric}, we can also show that providers with different coverages will differentiate and segment the market. More providers will further lower the equilibrium prices (below monopoly price).}

\vspace{-5pt}
\subsection{Impact of Local Competition on Global WiFi}\label{subsec:symm-Skype}
Now we discuss the entry of the global provider into a competitive market $i$ with two providers, and evaluate the impact of competition on the global provider's entry. {We will also consider the \emph{local monopoly benchmark}, where a monopoly local provider has a coverage of $G_{i,1}+G_{i,2}$, in which case the global provider's decision has been discussed in Section~\ref{subsec:SkypeUsage} by assuming $G_i=G_{i,1}+G_{i,2}$.} {Since any existing local provider's coverage is relatively small and their WiFi hotspots are often deployed at different locations, we assume the aggregate coverage $G_i\leq 1$.} 

To {analytically characterize the pricing decisions with the global provider's market entry and the impact of local competition},  we focus on the symmetric coverage setting, i.e., providers 1 and 2 each covers $G_{i}/2$.\footnote{{The asymmetric coverage setting can be analyzed in a similar way. Compared to symmetric coverage case, in this case, price competition mitigates (with non-zero prices in Theroem~\ref{thm:asymmetric}) and the global provider's price should be higher (compared to Theorem \ref{thm:comp_price}). Different from Theorem \ref{thm:bargaincomp}, The global provider will also decide different revenue sharing portions with the two differentiated providers. One can view the asymmetric coverage case as a partial competition scenario between the monopoly in Section \ref{subsec:SkypeUsage} (without competition) and perfect competition in Section \ref{subsec:symm-Skype}.}}
According to Proposition~\ref{prop:symmetric}, all users are  served by providers 1 and 2 at the same zero price before the global provider enters. {Such a competition has the following positive and negative impacts on the global provider's market entry:}
\begin{itemize}
\item \emph{Integration of WiFi coverage {(see Theorem \ref{thm:comp_price})}:} By integrating the local providers' WiFi networks, the global provider is able to provide the best service quality in terms of WiFi coverage, {which is important to} high-valuation users.
\item \emph{Larger bargaining power {(see Theorem~\ref{thm:bargaincomp}}):} As local providers' equilibrium revenues are zero, they are more willing to collaborate with the global provider. {This means that the global provider may share less revenue with each of the local provider.}
\item \emph{Severe price competition {(see Theorem \ref{thm:comp_price})}:} The zero local WiFi price and full user subscription {make it difficult for} the global provider to charge a high price to the local users.
\end{itemize}

By using backward induction,
we have the following result about the global provider's equilibrium usage-based pricing decision in Stage I (by considering users' responses in Stage II).

\begin{theorem}\label{thm:comp_price}
{At the market entry equilibrium with two symmetric coverage local providers, the global WiFi usage-based price is always} in the low price regime (i.e., $p_i^{Glob}\leq \ln(2)/2$), 
which is less than the price in the {local  monopoly benchmark}  (see Theorem~\ref{thm:win_win} in Section~\ref{subsec:bargain}). {All local users with a $\theta\in[2p_i^{Glob}/\ln(2),1]$ will use the global WiFi service and demand the full usage level.}
\end{theorem}

The proof of Theorem~\ref{thm:comp_price} is given in Appendix~\ref{app:comp_price}. {Even with a lower price comparing to the local monopoly benchmark,}  the global provider can still make a profit by providing a better WiFi service in doubling local WiFi coverage and attract high-end users.

Next we derive the revenues for the global provider and both local providers. To serve local users and travelers, the global provider needs to share $\eta_{i,j}\in(0,1)$ to provider $j\in\{1,2\}$.
Due to the symmetric WiFi coverage,
 we will focus on the symmetric sharing case where  $\eta_{i,1}=\eta_{i,2}=\eta_i$. Recall {from Theorem~\ref{thm:comp_price}} that the global provider will  serve local users with $\theta\in[2p_i^{Glob}/\ln(2),1]$. Similar to (\ref{eq:SkypeBenefit}), (\ref{eq:Provider_i_Benefit}), and the global market decomposition in Section~\ref{subsec:bargain}, we can derive the revenue increases of the global provider and local provider $j\in\{1,2\}$ in market $i$ as the summation of revenue increases in serving local users and travelers, i.e., 
\begin{align}\label{eq:Skype_benefit_comp}
\!\!\!\!\!\!&\Delta \pi^{Glob}(\eta_i,p_i^{Glob})=(1-2\eta_i)G_ip_i^{Glob}\nonumber\\
\!\!\!\!\!\!&\cdot\bigg(N_i\bigg(1-\frac{2p_i^{Glob}}{\ln(2)}\bigg)
+\sum_{ m\neq i}\alpha_m^iN_m\bigg(1-\frac{3p_i^{Glob}}{2}\bigg) \bigg),
\end{align}
and for $j\in\{1,2\}$
\begin{align}\label{eq:provider_benefit_comp}
&\Delta \pi^{i,j}(\eta_i,p_i^{Glob})=\eta_iG_ip_i^{Glob}\nonumber\\
&\cdot \bigg(N_i\bigg(1-\frac{2p_i^{Glob}}{\ln(2)}\bigg)
+\sum_{ m\neq i}\alpha_m^iN_m\bigg(1-\frac{3p_i^{Glob}}{2}\bigg) \bigg).
\end{align}
Notice that both the global provider and the two local providers' {revenue increases} are  increasing in the number of travelers from other markets $\sum_{m\neq i}\alpha_m^i N_m$. According to \cite{harsanyi1986rational}, the generalized Nash (group) bargaining in market $i$ involving  three parties {can be formulated as the} following joint optimization problem:
\begin{align}\label{opt:comp_skype}
&\max_{\eta_i, p_i^{Glob}} \Delta\pi^{Glob}\Delta\pi^{i,1} \Delta\pi^{i,2},\nonumber\\
&\text{subject \ to,} \quad  0\leq \eta_i \leq 1,\nonumber\\
&\quad \quad \quad \  0\leq p_i^{Glob}\leq \ln(2)/2.
\end{align}
This above problem formulation assumes that the global provider and the two local providers would engage in the bargaining {simultaneously}. {In {\cite{OurReport}}, we also show that all the results derived from solving Problem (\ref{opt:comp_skype}) still hold {even when} the global provider bargains with each provider individually and simultaneously.} 

Notice that the objective function in Problem (\ref{opt:comp_skype}) can be rewritten as
\begin{align}\label{eq:DeltaDeltaDelta}
&\Delta\pi^{Glob}\Delta\pi^{i,1} \Delta\pi^{i,2}=\eta_i^2(1-2\eta_i)\bigg(G_ip_i^{Glob}\nonumber\\
&\ \cdot\bigg(N_i\bigg(1-\frac{2p_i^{Glob}}{\ln(2)}\bigg)
+\sum_{ m\neq i}\alpha_m^iN_m\bigg(1-\frac{3p_i^{Glob}}{2}\bigg) \bigg)\bigg)^3,\nonumber
\end{align}
which is {quasi-concave} in $\eta_i$, as it is increasing in $\eta_i\in[0,1/3)$ and decreasing in $\eta_i\in(1/3,1]$ for any fixed value of $p_i^{Glob}$. Then we have the following result.

\begin{theorem}\label{thm:bargaincomp}
{At the market entry equilibrium with two symmetric coverage local providers,} the global provider equally shares the revenue {among the two providers and himself} (i.e., $\eta_i^*=1/3$).
\end{theorem}

Recall that in a monopoly local market, the global provider needs to {give up most of his}  revenue {to the monopoly local provider, in order to compensate the local provider's loss of market share and revenue} due to the global provider's competition (see Proposition~\ref{prop:eta}). However, with severe competition, the two operators' revenues are already zero and cannot be further reduced after the global provider's entry. {Hence the global provider can decide a smaller revenue sharing ratio with each of the local provider.} {However, the total revenue ratio shared by the global provider to both providers ($2\eta^{\ast}_{i}=2/3$) can be larger than the revenue sharing in the local monopoly benchmark (see Fig.~\ref{fig:Skype_Provideri_ratio}).}
\del{Is the last sentence the correct understanding? If yes, then this needs to be added to the abstract, intro, conclusion and other proper places to make the discussions complete. (In fact, if the analysis can be generalized to the case of oligopoly, then the total revenue shared by the global provider to all local providers will be $(Z-1)/Z$ where $Z$ is the number of local providers in the same market?) Lingjie: I suggest we skip this guess as it takes much space to prove and adds little clean value. Still I add footnote 18 to show the oligopoly case.}

After determining $\eta_i^*=1/3$, Problem (\ref{opt:comp_skype}) becomes a single-variable optimization problem and can be easily solved through an efficient one-dimensional exhaustive search algorithm. Next we provide some interesting numerical results.

\begin{figure}[tt]
\centering\vspace{-5pt}
\includegraphics[width=0.4\textwidth]{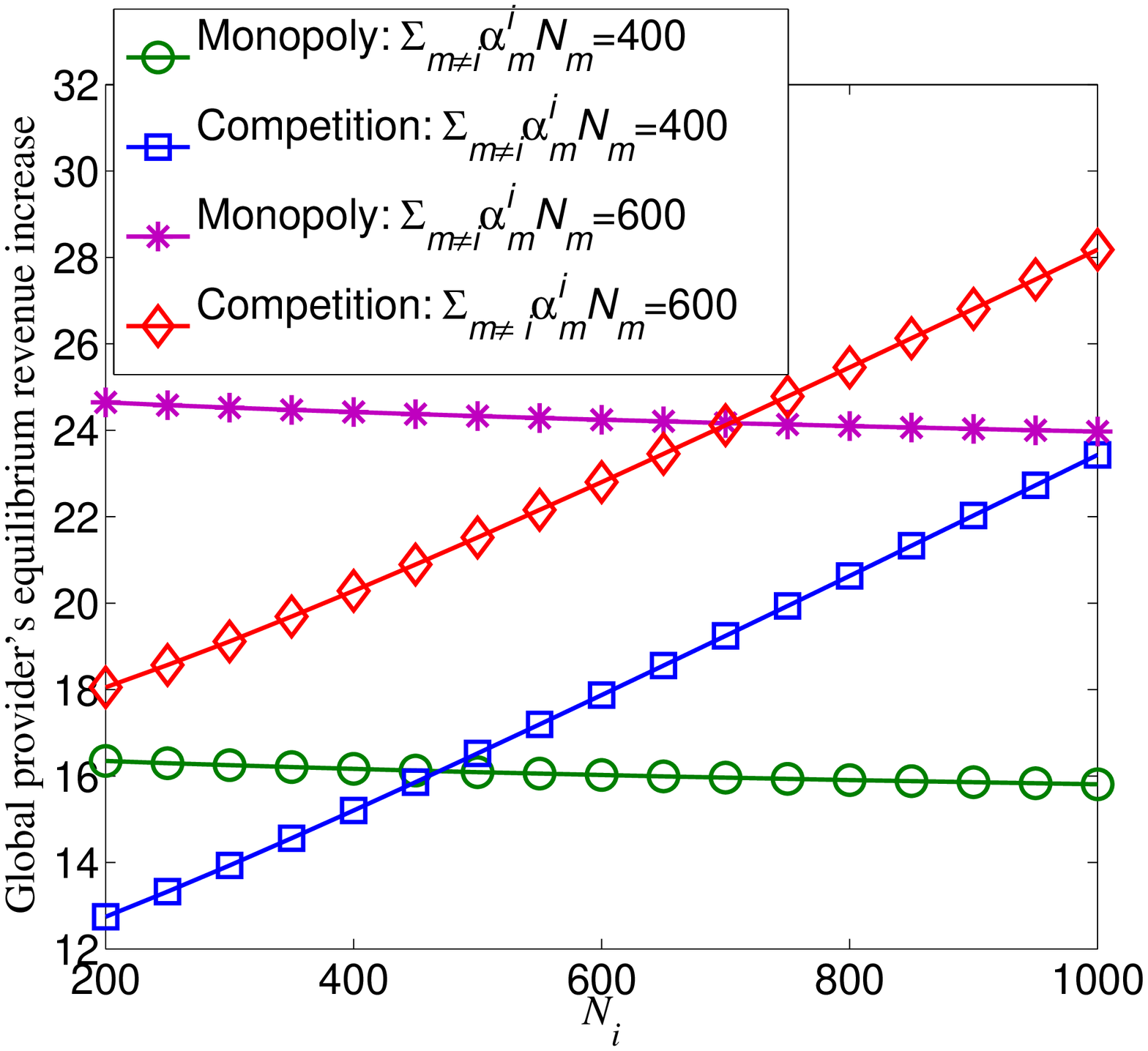}\vspace{-5pt}
\caption{The global provider's equilibrium revenue increases in monopoly and competition local markets as a function of traveler population $\sum_{m\neq i}\alpha_m^iN_m$ and local user population $N_i$. \del{need to change $\sum_{ j\neq i}\alpha_j^iN_j$ to $\sum_{m\neq i}\alpha_m^i N_m$ in the figure. Lingjie: Done thorughout all content.}}\vspace{-7pt} \label{fig:Monopoly_Comp_Skype}
\end{figure}

\begin{observation}\label{ob:bargaincomp}
The equilibrium revenue increases of the global provider and the local providers are increasing in $G_i$, $N_i$ and $\sum_{m\neq i}\alpha_m^i N_m$ (see Fig.~\ref{fig:Monopoly_Comp_Skype} for the global provider's case).
Compared to the local monopoly benchmark, {the global provider obtains a larger revenue increase in the competition case when there are a large number of local users $N_i$ ({hence there are a lot of high-valuation users to attract}), e.g., $N_{i}\geq 500$ when $\sum_{m\neq i}\alpha_m^iN_m=400$. Such a benefit diminishes when there are more travelers, e.g., the threshold of $N_{i}$ increases from 500 to 700 when  $\sum_{m\neq i}\alpha_m^iN_m$ changes from 400 to 600, meaning fewer travelers allow the global provider to better ``exploit'' the benefit of local competition.} 
\end{observation}

Intuitively, the local competition enables the global provider to easily enter and dominate the market including local users; however, the global provider can only charge a low price due to local competition and he also needs to share revenue with more providers. Compared to monopoly scenario, the global provider's revenue may decrease when local user number does not dominate traveler number.

{Note that the analysis in this subsection can be extended to oligopoly case, by considering $M$ local providers in market $i$ with identical coverage $G_i/M$. We can still show the low price regime holds for the global WiFi service, as in Theorem~\ref{thm:comp_price}. Following this, the Nash bargaining in (\ref{opt:comp_skype}) extends to have $M$ rather than 2 local revenue increases in the objective, where the global provider shares $\eta_i$ revenue with each provider and leaves $1-M\eta_i$ protion for himself. Similar to Theorem~\ref{thm:bargaincomp}, the bargaining outcome is $\eta_i^*=1/(M+1)$.}

\section{Conclusion}
{{Our study in this paper is motivated by the different pricing practices of local WiFi providers such as AT\&T in USA and PCCW in Hong Kong as well as a new kind of global WiFi provider represented by Skype WiFi.} 
We first show that in a local market, given abundant WiFi capacity, flat-rate pricing leads to higher revenue for a local provider than usage-based pricing. 
We {further} study how a global WiFi provider (e.g., Skype) cooperates with many local providers in using their WiFi infrastructures to provide a global WiFi service.
We explain why the global provider adopts usage-based pricing and gives away the majority of his revenue to the local provider. 

{There are some possible directions to extend our results. First, we can consider a user's traffic model is inelastic rather than elastic in the current model. For example, we can use a Sigmoid function to model a user's utility in using video conferencing, and the analysis will be more complicated. Second, this paper focuses on the steady state about users' service choices, and it is also interesting to study users' dynamics in network selection during the adoption process (as in \cite{Niyato1,Niyato2}). Third, we can further investigate the cellular data servcies' possible competition with local WiFi services. For example, we can introduce a reservation payoff for each user, and a user will choose the WiFi service only if his payoff is larger than this reservation payoff. Our main results should hold, and a larger reservation payoff encourages the WiFi provider to charge a smaller WiFi price to attract the subscribers. Finally, one may study how the network congestion affects the competition between two local providers. Different from Section 7.1, we believe that the congestion can help differentiate the service qualities of different providers, and even with the same coverage, it is unlikely to result in current non-profitable outcome (as in Proposition 5). One provider caters to the high-type users while the other covers many low-type users.}

\appendices
\section{Proof of Proposition~\ref{prop:StageII}}\label{app:StageII}
The proof consists three parts for three price regimes.
\subsection{{The Low Price Regime $p_i^{Glob}\leq {\ln(2)}/{2}$}}
In this regime, from the global provider's local users with types $\theta\in[p_i^{Glob},2p_i^{Glob})$ will demand partial usage levels (i.e., $d^*(\theta,p_i^{Glob})<1$) and users with types $\theta\in[2p_i^{Glob},1]$ will demand the full usage level (i.e., $d^*(\theta,p_i^{Glob})=1$). 

{Next we show that no user will choose the local provider $i$'s service i this regime.}
We can classify local users into three classes {based on $\theta$}.
\begin{itemize}
\item \emph{Local users with  $\theta\in[p_i^{Glob},{1}/{2}]$:} 
{these users will obtain {non-positive} payoffs (out of service) if choosing local provider $i$, but can obtain positive payoffs by choosing global WiFi.}
\item \emph{Local users with  $\theta\in[{1}/{2},2p_i^{Glob})$:} {a user can obtain the following maximum payoff if choosing global WiFi,}
%
    \begin{align}
    &u_i(\theta,p_i^{Glob},d^*(\theta,p_i^{Glob}))\nonumber\\ &\quad \quad \quad =G_i\left(\theta\ln\left(\frac{\theta}{p_i^{Glob}}\right)-\theta+p_i^{Glob}\right).\nonumber
    \end{align}
   {The same user's maximum payoff if choosing the local provider} $i$ is
    \[
    u_i(\theta,P_i^*,1)=G_i\left(\theta\ln(2)-\frac{\ln(2)}{2}\right).
    \]
    The payoff improvement of switching to global WiFi 
    \begin{align}\label{eq:Delta_u}
    \Delta u_i(\theta,p_i^{Glob})=& G_i\bigg(\theta\ln\left(\frac{\theta}{p_i^{Glob}}\right)-\theta+p_i^{Glob}\nonumber \\ &-\left(\theta-\frac{1}{2}\right)\ln(2)\bigg).
    \end{align}
   {Since $\theta<2p_i^{Glob}$, hence 
    \[
    \frac{\partial\Delta u_i(\theta,p_i^{Glob})}{\partial \theta}=G_i\ln\left(\frac{\theta}{2p_i^{Glob}}\right)<0,
    \]
    which shows the monotonicity of $\Delta u_i(\theta,p_i^{Glob})$ in $\theta$. Additionally, we can show $\Delta u_i(1/2,p_i^{Glob})>0$ and $\Delta u_i(2p_i^{Glob},p_i^{Glob})\geq 0$. }Thus we conclude that all users with types $\theta\in[{1}/{2},2p_i^{Glob})$ benefit from switching from provider $i$'s service to the global WiFi.
\item \emph{Local users with  $\theta\in[2p_i^{Glob},1]$:} these users demand a full usage level {with the global WiFi}. A type-$\theta$ user's optimal payoff {if choosing global WiFi} is
    \[
    u_i(\theta,p_i^{Glob},1)=G_i(\theta\ln(2)-p_i^{Glob}),
    \]
    which is larger than $u_i(\theta,P_i^*,1)$ due to $p_i^{Glob}<{\ln(2)}/{2}$. Thus these users {are better off by choosing} global WiFi.
\end{itemize}
{This complete the proof for the low price regime.}

\subsection{ Medium Price Regime ${\ln(2)}/{2}<p_i^{Glob}<{1}/{2}$} 
{By checking (\ref{eq:Delta_u}), the highest type (i.e., $\theta=1$) user will stay with provider $i$'s service due to $p_i^{Glob}>{\ln(2)}/{2}$}, and at least users with types $\theta\in[p_i^{Glob},{1}/{2})$ will subscribe to global WiFi. {Due to the continuity of $\theta$, we can show that there exists a}  
partition threshold $\theta_i^{th}$, where the type-$\theta_{i}^{th}$ user is indifferent in choosing between the two services.

{Notice that if  a type-$\theta_i^{th}$ user chooses global WiFi, he will demand only a partial usage, which means $\theta_i^{th}<2p_i^{Glob}$.  Assume that this is not the case,} then we will have  $u_i(\theta_i^{th},p_i^{Glob},1)=u_i(\theta_i^{th},P_i^*,1)$, i.e.,
\[
G_i(\theta_i^{th}\ln(2)-p_i^{Glob})=G_i(\theta_i^{th}\ln(2)-\frac{\ln(2)}{2}),
\]
{which violates the assumption $p_i^{Glob}>{\ln(2)}/{2}$ in this medium price regime.} Thus any user with type $\theta\in[p_i^{Glob},\theta_i^{th}{)}$ will demand partial usage levels in global WiFi, and we have $\Delta u_i(\theta_i^{th},p_i^{Glob})=0$ which results in (\ref{eq:theta_th}). Since $2p_i^{Glob}>\theta_i^{th}$, we have
\[
\frac{\partial \Delta u_i(\theta_i^{th},p_i^{Glob})}{\partial \theta_i^{th}}=G_i\ln\left(\frac{\theta_i^{th}}{2p_i^{Glob}}\right)<0,
\]
and thus the solution to (\ref{eq:theta_th}) is unique. 

Next we examine the relationship between $\theta_i^{th}$ and $p_i^{Glob}$, which are the only two variables in (\ref{eq:theta_th}). By checking
\[
\frac{\partial \Delta u_i(\theta_i^{th},p_i^{Glob})}{\partial p_i^{Glob}}=g(M_i,B)\left(-\frac{1}{p_i^{Glob}}+1\right),
\]
we have the following relationship by using the implicit function theorem,
\begin{equation}
\frac{d\theta_i^{th}}{dp_i^{Glob}}=-\frac{{\partial \Delta u_i(\theta_i^{th},p_i^{Glob})}/{\partial p_i^{Glob}}}{{\partial \Delta u_i(\theta_i^{th},p_i^{Glob})}/{\partial \theta_i^{th}}}=\frac{\frac{1}{p_i^{Glob}}-1}{\ln\left(\frac{\theta_i^{th}}{2p_i^{Glob}}\right)}<0,
\end{equation}
which shows $\theta_i^{th}$ is decreasing in price $p_i^{Glob}$. The intuition is that as the price increases, fewer users will switch from provider $i$'s service to global WiFi, {the threshold  $\theta_i^{th}$ decreases, and the subscriber number of global WiFi decreases.}

\subsection{{High Price Regime $p_i^{Glob}>{1}/{2}$}}

{In this regime,  local users with types $\theta\in[{1}/{2},p_i^{Glob})$ will receive negative payoffs by switching to global WiFi, hence they will stay with provider $i$'s and  receive positive payoffs. }

{Next we prove that no local users with types $\theta\in[p_i^{Glob},1]$ will switch to global WiFi.} The high price {$p_i^{Glob}>{1}/{2}$} can only attract those users with types $\theta>p_i^{Glob}$, and even the highest type (i.e., $\theta=1$) user will not request full usage level (i.e., $d^*(1,p_i^{Glob})<1$). A type $\theta{>} p_i^{Glob}$ user's payoff improvement $\Delta u_i(\theta,p_i^{Glob})$ by switching to global WiFi is given in (\ref{eq:Delta_u}). {Since}
\[
\Delta u_i(\theta,p_i^{Glob})|_{\theta=p_i^{Glob}}=-\left(p_i^{Glob}-\frac{1}{2}\right)\ln(2)<0,
\]
and 
\[
\frac{\partial \Delta u_i(\theta,p_i^{Glob})}{\partial \theta}=\ln\left(\frac{\theta}{2p_i^{Glob}}\right)<\ln\left(\theta\right)\leq 0
\]
due to $p_i^{Glob}>{1}/{2}$ and $\theta\leq 1$, we conclude that the payoff improvements of all users with types $\theta\in[p_i^{Glob},1]$ are negative by switching to global WiFi. Thus they will stay with provider $i$'s service.  Global WiFi has no users locally.

{
\section{Proof of Theorem~\ref{thm:win_win}}\label{app:lowprice}
To prove Theorem~\ref{thm:win_win}, we will first analyze the equilibrium of Stage I in high and low price regimes. Then we will compare it to the medium price regime, and show that both provider $i$ and the global provider prefer the equilibrium in the medium price regime. 

\subsection{Equilibrium Analysis of Stage I in the High Price Regime}

In the high price regime (i.e., $p_i^{Glob}>{1}/{2}$), provider $i$ is the monopolist to serve local users and the global provider is the monopolist to serve travelers in market $i$. Due to the high global WiFi price, all travelers will demand partial usage levels are in Fig.~\ref{fig:high}, and only travelers with types $\theta>p_i^{Glob}$ will subscribe to global WiFi. A type-$\theta>p_i^{Glob}$ traveler's demand is ${\theta}/{p_i^{Glob}}-1$. Similar to the derivation of (\ref{eq:usage_high}) in the high price regime in Section~\ref{subsec:LocalUsage}, we can show that the global provider's revenue increment is
\begin{align}
\Delta \pi^{Glob}(\eta_i,p_i^{Glob})=&(1-\eta_i)\sum_{ j\neq i}N_j\alpha_j^iG_ip_i^{Glob}\nonumber
\\ &\times\int_{p_i^{Glob}}^1\left(\frac{\theta}{p_i^{Glob}}-1\right)d\theta,
\end{align}
and its first partial derivative of $p_i^{Glob}$ is
\[
\frac{\partial \Delta \pi^{Glob}(\eta_i,p_i^{Glob})}{\partial p_i^{Glob}}=(1-\eta_i)\sum_{ j\neq i}N_j\alpha_j^iG_i({p_i^{Glob}-1})<0.
\]
Thus the global provider wants to decrease his price to the lowerbound ${1}/{2}$ of the high price regime to increase his revenue increment. Meanwhile, provider $i$'s revenue increment is
\begin{align}
\Delta \pi^i(\eta_i,p_i^{Glob})=&\frac{\ln(2)}{4}N_iG_i+\eta_i\sum_{ j\neq i}N_j\alpha_j^iG_ip_i^{Glob}\nonumber\\
&\times\int_{p_i^{Glob}}^1\left(\frac{\theta}{p_i^{Glob}}-1\right)d\theta,
\end{align}
which is also decreasing in $p_i^{Glob}$. Thus both the global provider and provider $i$ will agree on the price of $p_i^{Glob}={1}/{2}$ is achieved at the Nash bargaining. 

However, price ${1}/{2}$ is just a special case of the medium price regime, and it cannot outperform the equilibrium price in medium price regime. 

\subsection{Equilibrium Analysis of Stage I in the Low Price Regime}
In the low price regime (i.e., $p_i^{Glob}\leq {\ln(2)}/{2}$), no users choose provider $i$'s service and $\theta_i^{th}=1$. By following a similar analysis as in Section~\ref{subsec:SkypeUsage}, we can derive the revenue increment of the global provider as
\begin{align}
\Delta \pi^{Glob}(\eta_i,p_i^{Glob})=&(1-\eta_i)G_i(N_i+\sum_{ j\neq i}\alpha_j^iN_j)\nonumber\\ &\times (p_i^{Glob}-\frac{3}{2}(p_i^{Glob})^2),
\end{align}
as the global provider is a monopolist in market $i$. The revenue increment of provider $i$ is
\begin{align}
\Delta \pi^{i}(\eta_i,p_i^{Glob})=& G_i\bigg(-\frac{\ln(2)}{4}N_i+\eta_i(N_i+\sum_{ j\neq i}\alpha_j^iN_j)\nonumber\\ &\times (p_i^{Glob}-\frac{3}{2}(p_i^{Glob})^2)\bigg),
\end{align}
who loses all of his original revenue after the global provider's entry and hence needs to be compensated by the global provider in order to share the network. The optimization problem to derive Nash bargaining is
\begin{align}\label{eq:prob_Skype_Provider_i_lowprice}
\max_{\eta_i,p_i^{Glob}}&\ \Delta \pi^{Glob}(\eta_i,p_i^{Glob})\Delta\pi^{i}(\eta_i,p_i^{Glob})\nonumber\\
\text{subject\ to,\ } &\ \ \ \ \ \ \ \ \ 0<\eta<1, \nonumber\\
&\ \ \ \  \quad p_i^{Glob}\leq \frac{\ln(2)}{2}.
\end{align}
Notice that the two decision variables $p_i^{Glob}$ and $\eta_i$ {are not coupled with each other in the constraints of} Problem~(\ref{eq:prob_Skype_Provider_i_lowprice}), and both $\Delta \pi^{Glob}(\eta_i,p_i^{Glob})$ and $\Delta\pi^{i}(\eta_i,p_i^{Glob})$ are maximized at the monopoly usage-based price ${1}/{3}$ (see (\ref{eq:usage_price}) with $k=1$). Thus the optimal $p_i^{Glob*}$ is ${1}/{3}$ which is smaller than ${\ln(2)}/{2}$.

Now we will maximize the product $\Delta \pi^{Glob}(\eta_i,{1}/{3})\Delta \pi^{i}(\eta_i,{1}/{3})$ by choosing $\eta_i${, which can be shown concave in $\eta_i$.} 
By checking its first order condition, we can derive the optimal sharing ratio as
\begin{equation}
\eta_i^*=\max\left(\frac{\frac{3\ln(2)}{2}N_i}{N_i+\sum_{ j\neq i}\alpha_j^iN_j},\frac{1}{2}+\frac{\frac{3\ln(2)}{4}N_i}{N_i+\sum_{ j\neq i}\alpha_j^iN_j}\right),
\end{equation}
which is larger than ${1}/{2}$. This means that the global provider needs to share the majority of his revenue increment with provider $i$ as in Proposition~\ref{prop:eta}.

{Finally, we can compare the maximum bargaining objective $\Delta \pi^{Glob}(\eta_i^*,{1}/{3})\Delta \pi^{i}(\eta_i^*,{1}/{3})$ here to that in the medium price regime (i.e., (\ref{eq:prob_Skype_Provider_i_simple})). Figure~\ref{fig:LowMediumPrice_ProductRatio} shows the ratio of bargaining objectives between the low and medium price regimes as a function of traveler number $\sum_{ j\neq i}\alpha_j^iN_j$ and local user number $N_i$. The ratio is less than 1 in most scenarios and is larger than 1 only when $N_i$ is not comparable to $\sum_{j\neq i}\alpha_j^i N_j$ (e.g., when $N_i=350$ and $\sum_{j\neq i}\alpha_j^i N_j\geq 900$). Thus we can conclude that the bargaining objective in medium price regime is larger than that in low price regime, as long as we do not face the extreme case where $N_i$ is trivial compared to $\sum_{j\neq i}\alpha_jN_j$. This means that the optimal decision of $p_i^{Glob}$ in the medium price regime outperforms that in the low price regime under the practical parameter setting.}


\begin{figure}[tt]
\centering
\includegraphics[width=0.45\textwidth]{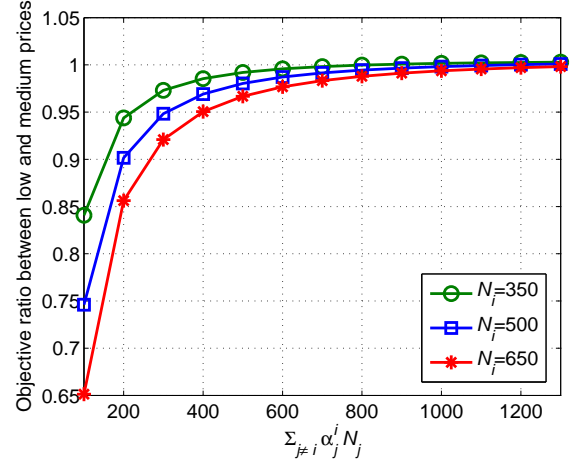}
\caption{The ratio of revenue increment product between low and medium price regimes as a function of traveler number $\sum_{ j\neq i}\alpha_j^iN_j$ and local user number $N_i$.} \label{fig:LowMediumPrice_ProductRatio}
\end{figure}

%

}

\section{Proof of Lemma~\ref{lem:relationship}}\label{app:relationship}

{Relationship (\ref{eq:relation_competition}) involves three thresholds of $\theta$.
A user with  $\theta=\frac{P_{i,j}}{G_{i,j}\ln(2)}$ is indifferent between not using the WiFi service and subscribing to provider $j$'s service, \emph{if provider $j$ is the monopolist in the market}. A user with $\theta=\frac{P_{i,1}-P_{i,2}}{(G_{i,1}-G_{i,2})\ln(2)}$ is indifferent between subscribing to any one of the two providers.} 
{These three thresholds must lie in $(0,1)$ at the equilibrium, otherwise at least one provider will have no subscribers and he will have an incentive to lower his price.}

{Given three thresholds, there are a total of six possible relationships in terms of their relative values. However,}  
%
we can show that $\frac{P_{i,1}-P_{i,2}}{(G_{i,1}-G_{i,2})\ln(2)}<\frac{P_{i,1}}{G_{i,1}\ln(2)}$ is equivalent to $\frac{P_{i,1}}{G_{i,1}\ln(2)}<\frac{P_{i,2}}{G_{i,2}\ln(2)}$, hence either (\ref{eq:relation_competition}) or the following is true:
\begin{align}\label{eq:proof_comp_12}\frac{P_{i,1}-P_{i,2}}{(G_{i,1}-G_{i,2})\ln(2)}<\frac{P_{i,1}}{G_{i,1}\ln(2)}<\frac{P_{i,2}}{G_{i,2}\ln(2)}.
\end{align}
In (\ref{eq:proof_comp_12}), however, users with $\theta>\frac{P_{i,1}-P_{i,2}}{(G_{i,1}-G_{i,2})\ln(2)}$ are only interested in provider 1's service, and provider 2 has no subscribers. Hence provider 2 has an incentive to decrease $P_{i,2}$ such that $\frac{P_{i,1}-P_{i,2}}{(G_{i,1}-G_{i,2})\ln(2)}>\frac{P_{i,2}}{G_{i,2}\ln(2)}$. {Thus  (\ref{eq:proof_comp_12}) can not be true at the equilibrium. This proves the Lemma.} 

\section{Proof of Theorem~\ref{thm:comp_price}}\label{app:comp_price}
 
We first show that no equilibrium exists in the high or the medium price regime, {and then discuss the low price regime.} 

We first look at the high price regime ($p_i^{Glob}>1/2$). 
{In this regime,} 
a type-$\theta$ user's payoff by choosing any local provider's service is $\theta{G_i}\ln(2)/2$, and his payoff by choosing global WiFi is $G_i(\theta\ln(\theta/p_i^{Glob})-\theta+p_i^{Glob})$ (based on (\ref{eq:valuation_basis}) and (\ref{eq:usage_demand})). {Notice that  a local user with a $\theta< p_i^{Glob}$ will never choose the global WiFi }{, since in that case the user's demand will be zero according to (\ref{eq:usage_demand})}. {Hence we only need to consider local users with $\theta\in [p_i^{Glob},1]$.} 
{For a user with $\theta\in [p_i^{Glob},1]$,} he {will only demand a partial usage level and} his payoff improvement  by switching from local WiFi to global WiFi is
\begin{equation}\label{eq:Deltav}
\Delta v^i(\theta,p_i^{Glob})=G_i\theta\left(\ln\left(\frac{\theta}{p_i^{Glob}}\right)-1+\frac{p_i^{Glob}}{\theta}-\frac{\ln(2)}{2}\right).
\end{equation}
The first partial derivative of (\ref{eq:Deltav}) over $\theta$ is 
\[
\frac{\partial \Delta v^i (\theta,p_i^{Glob})}{\partial \theta}=G_i\ln\left(\frac{\theta}{\sqrt{2}p_i^{Glob}}\right),
\] 
{which is negative if  $\theta \in [p_i^{Glob}, \sqrt{2}p_i^{Glob})$, is zero if  $\theta = \sqrt{2}p_i^{Glob}$, and is positive if $\theta \in(\sqrt{2}p_i^{Glob},1]$.} Thus (\ref{eq:Deltav}) reaches its minimum at $\theta=\sqrt{2}p_i^{Glob}$, and {reaches} its maximum either at $\theta=p_i^{Glob}$ or $\theta=1$. We can show that 
\[
\Delta v^i(\theta,p_i^{Glob})\mid_{\theta=p_i^{Glob}}=-G_ip_i^{Glob}\ln(2)/2<0\] 
and 
\begin{align}
&\Delta v^i(\theta,p_i^{Glob})\mid_{\theta=1}\nonumber\\
&=G_i\left(-\ln(p_i^{Glob})-1+p_i^{Glob}-{\ln(2)}/{2}\right)<0.\nonumber\end{align} 
Hence $\Delta v^i(\theta,p_i^{Glob})<0$ for any $\theta\in[p_i^{Glob},1]$, and no users will choose global WiFi in Stage II due to his high price. {This means that the global provider's revenue is zero, and he has an incentive to reduce his price in Stage I to improve the revenue. Hence an equilibrium can not exist in this high price regime.}

Next we look at the medium price regime ($\ln(2)/2<p_i^{Glob}\leq 1/2$).
For a type $\theta>p_i^{Glob}$ user, {he may demand a partial or full usage level and} his payoff improvement by switching from any local provider's service to global WiFi is 
\begin{align}\label{eq:Delta_v_medium}
&\Delta v^i(\theta,p_i^{Glob})=\nonumber\\
&\begin{cases} G_i\theta\left(\ln\left(\frac{\theta}{p_i^{Glob}}\right)-1+\frac{p_i^{Glob}}{\theta}-\frac{\ln(2)}{2}\right), \text{if}\ {\frac{\theta}{p_i^{Glob}}\in[1,2)},\\ G_i \theta\left(\frac{\ln(2)}{2}-\frac{p_i^{Glob}}{\theta}\right), \text{if} \  {\theta\in[2p_i^{Glob},1]} .\end{cases}
\end{align}
 Similar to our previous analysis in the high price regime, we can show that $\Delta v^i(\theta,p_i^{Glob})$ in (\ref{eq:Delta_v_medium}) is decreasing in $\theta\in[p_i^{Glob},\sqrt{2}p_i^{Glob}]$ and increasing in $\theta\in(\sqrt{2}p_i^{Glob},1]$. Thus $\Delta v^i(\theta,p_i^{Glob})$ reaches its maximum at either $\theta=p_i^{Glob}$ or $\theta=1$. By checking $\Delta v^i(\theta,p_i^{Glob})\mid_{\theta=p_i^{Glob}}<0$ and 
$\Delta v^i(\theta,p_i^{Glob})\mid_{\theta=1}=G_i(\ln(2)/2-p_{i}^{Glob})<0$  in this medium price regime, 
we conclude that $\Delta v^i(\theta,p_i^{Glob})<0$ for any $\theta\in[p_i^{Glob},1]$. 
{This means that the global provider's revenue is zero, and he has an incentive to reduce his price in Stage I to improve the revenue. Hence an equilibrium can not exist in this medium price regime.}


Finally we are ready to examine users' choices in the low price regime ($p_i^{Glob}\leq \ln(2)/2$). The analysis of this regime is similar to the {previous two regimes, where we will study how the payoff improvement $\Delta v^i(\theta,p_i^{Glob})$ changes with $\theta$.} We can show that $\Delta v^i(\theta,p_i^{Glob})\mid_{\theta=1}\geq 0$ (since $p_i^{Glob}\leq \ln(2)/2$), $\Delta v^i(\theta,p_i^{Glob})\mid_{\theta=2p_i^{Glob}}=G_ip_i^{Glob}(\ln(2)-1)<0$, and $\Delta v^i(\theta,p_i^{Glob})$ is {strictly} increasing in $\theta\in[2p_i^{Glob},1]$, we conclude that there exists a unique threshold $\tilde{\theta}_{th}$ such that all users with $\theta\in {(}\tilde{\theta}_{th},1]$ will choose global WiFi rather than any local WiFi service. {A} user with $\theta=\tilde{\theta}_{th}$ is indifferent in choosing any service since his payoff improvement is zero (i.e., 
$\Delta v^i(\theta,p_i^{Glob})\mid_{\theta=\tilde{\theta}_{th}}=G_i(\theta\ln(2)/2-p_i^{Glob})=0$). This means that $\tilde{\theta}_{th}=2p_i^{Glob}/\ln(2)$. As $\tilde{\theta}_{th}>2p_i^{Glob}$, the local users who choose global WiFi will demand a full usage level (i.e., $d^*(\theta,p_i^{Glob})=1$).

\end{document}